\documentclass[preprint,12pt]{elsarticle}
\usepackage[margin=1in]{geometry}

\usepackage{tabulary}
\usepackage{graphicx}
\usepackage{amssymb}

\journal{Journal Name}

\begin{document}

\begin{frontmatter}


\title{Data-driven analysis of thermal simulations, microstructure and mechanical properties of Inconel 718 thin walls deposited by metal additive manufacturing }



\author[1]{Lichao Fang}
\author[1]{Lin Cheng}
\author[2]{Jennifer A. Glerum}
\author[1,3]{Jennifer Bennett}
\author[1]{Jian Cao}
\author[1]{Gregory J. Wagner\corref{cor1}}

\cortext[cor1]{Corresponding author}

\address[1]{Department of Mechanical Engineering, Northwestern University, Evanston, IL, 60208, USA}
\address[2]{Department of Materials Science and Engineering, Northwestern University, Evanston, IL, 60208, USA}
\address[3]{DMG MORI, Hoffman Estates, IL 60192, USA}

\begin{abstract}
The extreme and repeated temperature variation during additive manufacturing of metal parts has a large effect on the resulting material microstructure and properties. The ability to accurately predict this temperature field in detail, and relate it quantitatively to structure and properties, is a key step in predicting part performance and optimizing process design. In this work, a finite element simulation of the Directed Energy Deposition (DED) process is used to predict the space- and time-dependent temperature field during the multi-layer build process for Inconel 718 walls. The thermal model is validated using the dynamic infrared (IR) images captured in situ during the DED builds, showing good agreement with experimental measurements. The relationship between predicted cooling rate, microstructural features, and mechanical properties is examined, and cooling rate alone is found to be insufficient in giving quantitative property predictions. Because machine learning offers an efficient way to identify important features from series data, we apply a 1D convolutional neural network (CNN) data-driven framework to automatically extract the dominant predictive features from simulated temperature history. The relationship between the CNN-extracted features and the mechanical properties is studied. To interpret how CNN performs in intermediate layers, we visualize the extracted features produced on each convolutional layer by a trained CNN. Our results show that the results predicted by the CNN agree well with experimental measurements and give insights for physical mechanisms of microstructure evolution and mechanical properties. 

\end{abstract}


\end{frontmatter}


\section{INTRODUCTION}
\label{S:1}

Metal additive manufacturing (AM) is a technology that can be used to build parts layer-by-layer, allowing the fabrication of parts with more complex geometry and reduced costs compared with traditional manufacturing techniques \cite{keller2017application, debroy2018additive}.  Directed energy deposition (DED) is one popular metal additive manufacturing process \cite{heigel2015thermo} in which metal powder is delivered by one or more nozzles \cite{wang2016effect}. A focused heat source, such as a laser, is used to melt the injected metal material locally. Parts are built progressively as each layer is scanned and melted in a predetermined pattern. 

During the DED process, parts undergo repetitive thermal heating and cooling cycles due to the deposition of multiple layers. The resulting complex thermal field in parts, both during and after solidification, has significant effects on the final material microstructure and mechanical properties, such as yield stress, yield strain, ultimate tensile strength, and failure stress \cite{gan2017modeling, wei2015evolution, manvatkar2011estimation}. However, it is time-consuming and expensive to conduct DED experiments to optimize process parameters and tool paths for a given geometry to yield parts with good mechanical properties. Computational models can be an efficient approach to obtaining temperature histories of parts, which can be related to the microstructure and mechanical properties.

To predict the thermal field, many researchers have used the finite element method to solve the heat equation and simulate the transient temperature field in AM.  For most DED thermal models, the boundary condition on the outer part surface assumes convection with a constant convection coefficient \cite{li2014parametric, foroozmehr2016finite, kelly2004microstructural, labudovic2003three, jendrzejewski2004temperature, wang2008residual, ghosh2005three, zekovic2005thermo, tan2018process}. However, DED processes typically include a forced shield gas flow, with a flow velocity that varies over the part surface; \citet{heigel2015thermo}, therefore, proposed a spatially varying convection coefficient model calibrated against measured thermocouple data, and showed a better match with experimental temperature histories compared with a uniform convection coefficient model. 

Calibrations of thermal DED models are also challenging. Almost all previous calibrated thermal models for multi-layer deposition have been based on thermocouple measurements taken far from the laser spot \cite{chukkan2015simulation, denlinger2016thermal, qian2005influence}. However, it is difficult to directly measure the temperature in or near the melt pool region with thermocouples because of the extreme temperature range and the constantly changing geometry. Alternatively, dynamic infrared (IR) images measured by IR cameras have been used to calibrate thermal models \cite{bai2015modeling, bai2013improving}. An IR camera can capture the emitted thermal radiation at the part surface, including near the melt pool, providing a complement to thermocouple data for calibrating and validating thermal models \cite{wolff2019experimentally, bennett2018cooling, yan2018review, bai2013improving, bai2015modeling, johnson2018simulation, price2012evaluations, price2013experimental}. For instance, \citet{bai2015modeling} captured IR images to verify a moving heat source thermal model for induction assisted weld-based additive manufacturing (WAM). \citet{yang2017thermal} also used IR imaging data to calibrate the thermal model for a single pass multi-layer gas metal arc welding (GMAW) process; this work only studied two layers of deposition rather than a full part-scale model. 

Nickel-based alloys, such as Inconel 718, have been widely used in AM applications (e.g., turbine blades and combustion chambers) because of their excellent tensile strength, high-temperature yield strength, creep properties, and corrosion resistance \cite{wei2015evolution, zhou2009effects, popovich2017functionally, thomas2006high, raghavan2016numerical}. Mechanical properties are found to depend not only on grain size and microstructure orientation, but also on the precipitate distribution of the material \cite{popovich2017functionally, manvatkar2011estimation}. For example, Laves phases are brittle precipitates in Inconel 718 usually formed during Nb segregation in the inter-dendrite regions in solidification. Laves phases can reduce the mechanical properties of the material, such as lowering the yield strength and Young's modulus for Inconel 718 \cite{popovich2017functionally}. The microstructure can be affected by the chemical composition and thermal conditions during the AM process \cite{thomas2006high}. Therefore, it is important to investigate the detailed effects of thermal history on microstructure and mechanical properties. 

Much previous work uses cooling rate to describe the relationship between temperature, microstructure, and properties. Researchers have found that cooling rate influences grain size, dendrite arm spacing, microsegregation, precipitate formation \cite{gan2019benchmark, bennett2018cooling, rappaz1993probabilistic}, anisotropy, porosity, and strength \cite{wolff2019experimentally}. High cooling rates can lead to faster solidification that results in finer microstructure and enhanced mechanical properties \cite{li2017melt}. However, far less previous work investigates whether cooling rate is the only important factor for microstructure and mechanical properties. It has been found that the combination of thermal gradients, cooling rates (the average rate of temperature change during solidification, in units of temperature per time), and solidification rates (the speed of the solidification front, in units of length per time) can influence the liquid-solid interface stability and then affect the dendrite arm spacing and final properties \cite{kurz1981dendrite, raghavan2016numerical, raghavan2013heat}. However, the complex thermal effects on microstructure and mechanical properties are not fully understood. This motivates us to develop a method to automatically extract features from temperature history and investigate the correlation between thermal history and mechanical properties.

Machine-learning techniques provide an effective way to extract information from a signal or time series. For example, a convolutional neural network (CNN) can learn local patterns from input data through convolutions without a priori feature selection \cite{herriott2020predicting, guo2016convolutional, kim2020prediction, fonda2019deep}. CNNs have been highly successful in many applications, such as speech recognition, self-driving vehicle control, and computer vision \cite{fonda2019deep, abdeljaber2017real, kiranyaz2015real}. Recently, one-dimensional (1D) CNNs have been used to extract features from 1D input data, such as a heart signal or other time-series data \cite{abdeljaber20181, abdeljaber2017real, kiranyaz2015real, kiranyaz20191d, avci2018wireless, yang2019fault}. CNNs can learn local features from raw data, and then extract more global and high-level features in deeper convolutional layers \cite{schirrmeister2017deep, abdeljaber20181, abdeljaber2017real}. Measured or simulated thermal history data from DED-built parts can be regarded as a dynamic time series. Recently, it was shown that a 2D CNN can be used to correlate the experimentally measured thermal history in AM with material properties \cite{xie2021mechanistic}. In the current study, we use 1D CNN to automatically extract features from {\em simulated} temperature history and investigate thermal effects on mechanical properties. Although features and correlations discovered by CNNs can be difficult to interpret \cite{schirrmeister2017deep, mahendran2016visualizing, yosinski2015understanding}, in this work, we visualize the intermediate convolutional layers to improve understanding of the underlying physics and investigate thermal effects on mechanical properties.  

In this work, a combination of the validated computational thermal model, together with a data-driven method based on a 1D CNN, is developed to accurately simulate the process and use the entire time-dependent temperature curve to predict the ultimate tensile strength (UTS) throughout the final part. First, we use a thermal model based on the finite element method with spatially varying surface convection for the DED multi-layer build process of Inconel 718 material. Three cases with different domain sizes and dwell times are investigated. The thermal model is validated using IR images from in situ DED experiments. Microstructure characterization and tensile tests are also conducted for selected material samples. Then, a 1D CNN is used to extract features from simulated thermal histories to predict mechanical properties. The correlation between thermal history, microstructure, and mechanical properties is investigated and discussed. To better understand the physical mechanism, we also visualize the extracted features from intermediate convolutional layers to interpret the correlation between thermal histories and properties. This work demonstrates an efficient way to simulate the thermal history of DED deposition, predict properties from such a thermal history by 1D CNN, and further understanding of thermal effects on solidification and mechanical properties.

\section{RESULTS}
\subsection{Thermal simulation and validation}
\begin{figure}
	\centering
	{
		\begin{minipage}[b]{3.1in}
			\centerline{\includegraphics[height=2in]{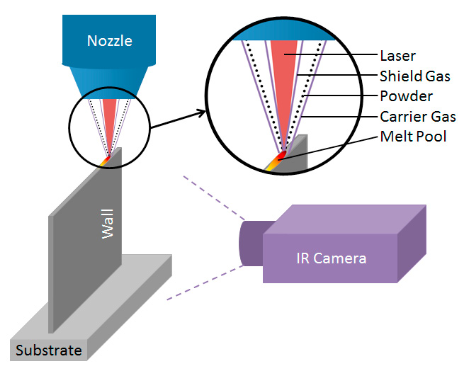}}	
			\centerline{(a)}
		\end{minipage}
	}
	\hfill
	{
		\begin{minipage}[b]{3.1in}
			\centerline{\includegraphics[height=2in]{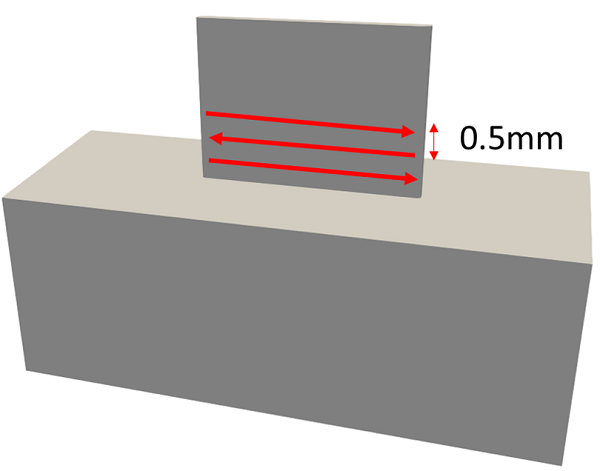}}	
			\centerline{(b)}
		\end{minipage}
	}
	\vfill
	{
		\begin{minipage}[b]{2in}
			\centerline{\includegraphics[height=2in]{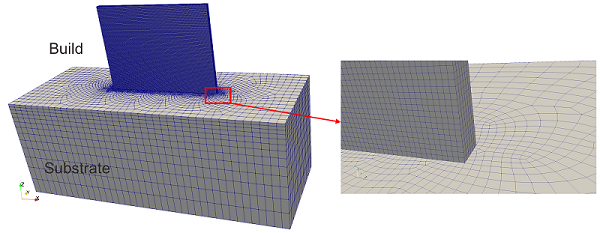}}	
			\centerline{(c)}
		\end{minipage}
	}
	\vfill
	{
		\begin{minipage}[b]{3.1in}
			\centerline{\includegraphics[height=1.7in]{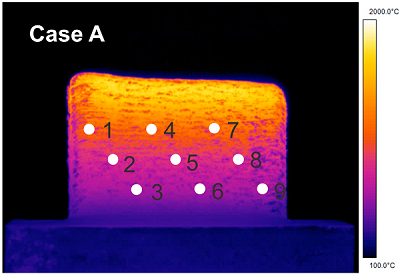}}	
			\centerline{(d)}
		\end{minipage}
	}
	\hfill
	{
		\begin{minipage}[b]{3.1in}
			\centerline{\includegraphics[height=1.7in]{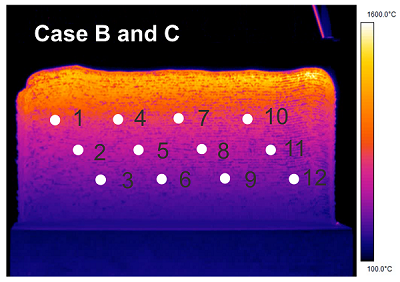}}	
			\centerline{(e)}
		\end{minipage}
	}
	\caption{Schematic of computational domain and experiment. (a) DED experiment process (reprinted from ref.\cite{bennett2018cooling} with permission from Elsevier). (b) Schematic of toolpath design during the DED process. (c) Geometry and mesh for computational domain. Case A is shown; meshes for Cases B and C are similar. The top part of the mesh is for thin wall build while the bottom part is substrate. The inset shows mesh details near the base of the wall. (d,e) Temperature images measured by the IR camera with locations of the coupon specimens for Case A, B and C. }
	\label{fig:meshtoolpath}
\end{figure}

In this work, three cases of thin wall, multi-layer depositions, during the DED process are simulated using the computational thermal model (see Methods section). The three cases are 80 mm walls with 0 s dwell time (Case A), 120 mm walls with 0 s dwell time (Case B), and 120 mm walls with a 5 s inter-layer dwell time (Case C). More details of the three cases are listed in Supplementary Table 1.  The schematic of the DED process is shown in Fig.~\ref{fig:meshtoolpath}a. The bidirectional tool path used in the simulation and experiment is shown in Fig.~\ref{fig:meshtoolpath}b. For the simulation, the geometry of the computational domain and meshes are shown in Fig.~\ref{fig:meshtoolpath}c. The top part of the domain is the thin Inconel 718 wall while the bottom part is the stainless steel 304 substrate. More process parameters and thermal-physical properties of material Inconel 718 and substrate material stainless steel 304 are listed in Supplementary Table 2, 3, and 4.

The model computes temperature as a function of time and space. Detailed temperature history is output at every time step for each specific probe point. To investigate the correlation between thermal history and mechanical properties, we choose the probe points in the simulation at the same locations as the experimental tensile test coupons. Further post-processing of the temperature solutions provides additional data used for analysis (see Methods). The proposed computational model can efficiently provide thermal history, cooling rate, and solidification rate to further investigate the correlation between thermal data, microstructure and mechanical properties. 

\begin{figure}
	\centering
	{
		\begin{minipage}[b]{2in}
			\centerline{\includegraphics[width=1.8in]{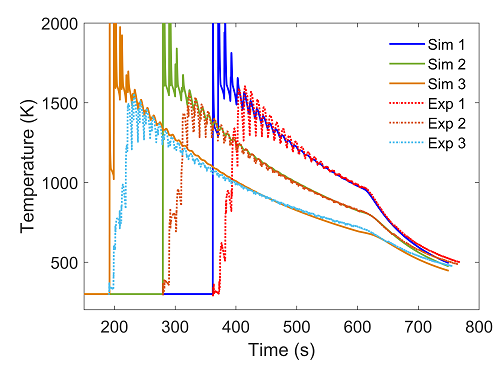}}	
			\centerline{(a)}
		\end{minipage}
	}
	\hfill
	{
		\begin{minipage}[b]{2in}
			\centerline{\includegraphics[width=1.8in]{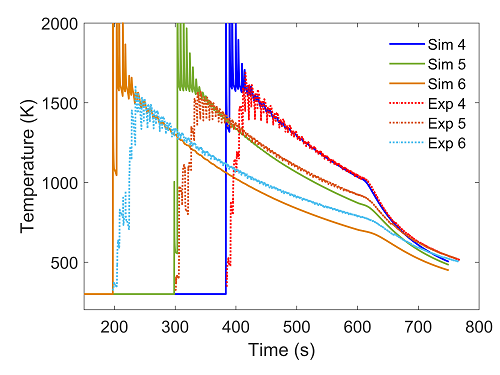}}	
			\centerline{(b)}
		\end{minipage}
	}
	\hfill
	{
		\begin{minipage}[b]{2in}
			\centerline{\includegraphics[width=1.8in]{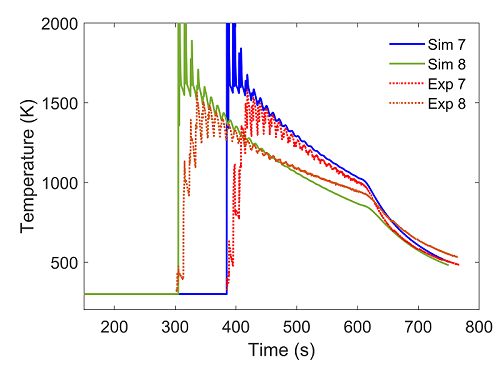}}	
			\centerline{(c)}
		\end{minipage}
	}
	\vfill
	{
		\begin{minipage}[b]{1in}
			\centerline{\includegraphics[width=1.8in]{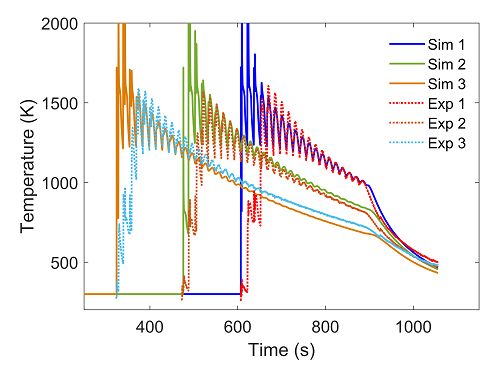}}	
			\centerline{(d)}
		\end{minipage}
	}
	\hfill
	{
		\begin{minipage}[b]{1in}
			\centerline{\includegraphics[width=1.8in]{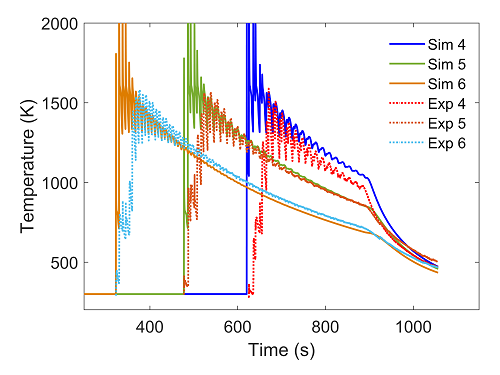}}	
			\centerline{(e)}
		\end{minipage}
	}
	\hfill
	{
		\begin{minipage}[b]{1in}
			\centerline{\includegraphics[width=1.8in]{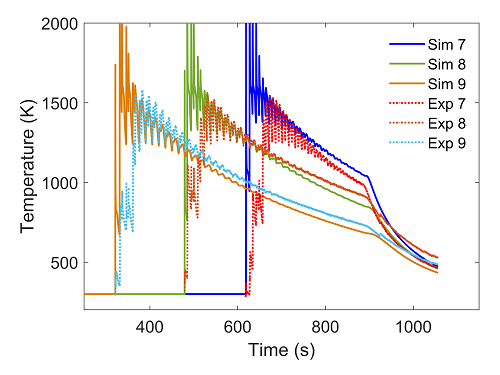}}	
			\centerline{(f)}
		\end{minipage}
	}
	\hfill
	{
		\begin{minipage}[b]{1in}
			\centerline{\includegraphics[width=1.8in]{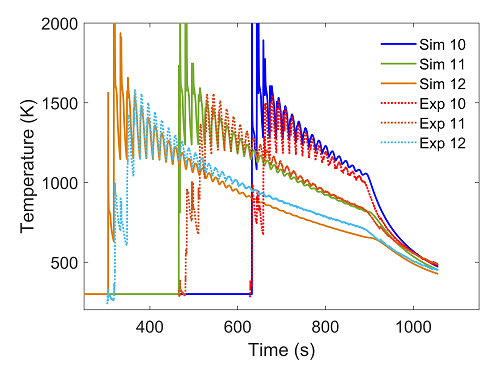}}	
			\centerline{(g)}
		\end{minipage}
	}
	\vfill
	{
		\begin{minipage}[b]{1in}
			\centerline{\includegraphics[width=1.8in]{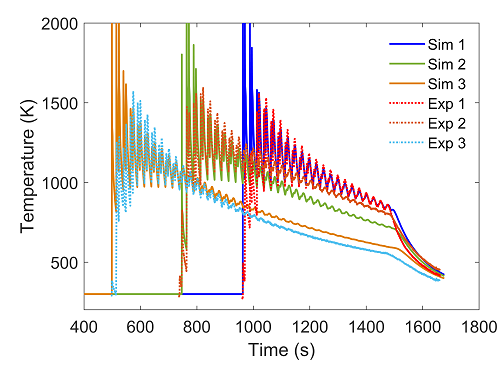}}	
			\centerline{(h)}
		\end{minipage}
	}
	\hfill
	{
		\begin{minipage}[b]{1in}
			\centerline{\includegraphics[width=1.8in]{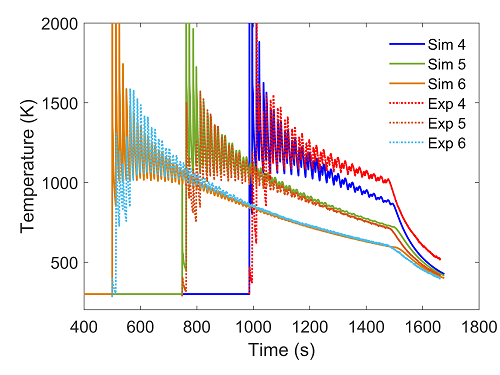}}	
			\centerline{(i)}
		\end{minipage}
	}
	\hfill
	{
		\begin{minipage}[b]{1in}
			\centerline{\includegraphics[width=1.8in]{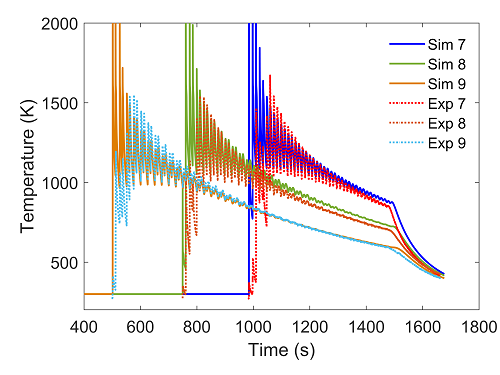}}	
			\centerline{(j)}
		\end{minipage}
	}
	\hfill
	{
		\begin{minipage}[b]{1in}
			\centerline{\includegraphics[width=1.8in]{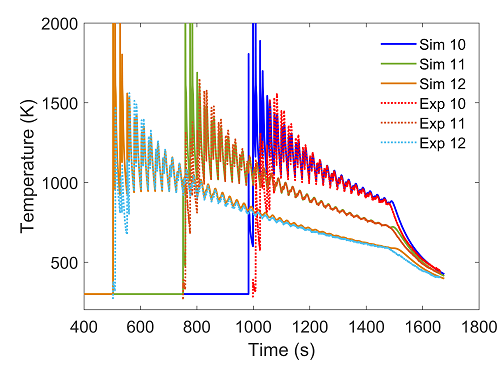}}
			\centerline{(k)}
		\end{minipage}
	}
	\caption{Comparison of simulated and experimental temperature histories. (a)-(c) Case A, 8 specimen locations. (d)-(g) Case B, 12 specimen locations. (h)-(k) Case C, 12 specimen locations. Point locations are labeled in Fig.~\ref{fig:meshtoolpath} (d) and (e)}
	\label{fig4.1:TemperatureHistory_all}
\end{figure}

The parameters in the spatially varied heat convection model, the laser absorption efficiency, and the emissivity values in the model are calibrated using the thermal history measured by the IR camera in the experiment. Fig.~\ref{fig4.1:TemperatureHistory_all} shows the temperature histories predicted by the calibrated computational models compared with experiments for Cases A, B, and C, respectively. The locations of point 1 through 8 for Case A and 1 through 12 for Cases B and C are labeled in Fig.~\ref{fig:meshtoolpath}d and e. 

The results show that the predicted thermal history agrees well with the experimental results for most specimen locations. For example, Fig.~\ref{fig4.1:TemperatureHistory_all}a shows a comparison of the thermal histories from simulation (solid lines) and experimental IR camera data (dashed lines) for points 1, 2, and 3 in Case A during the DED process. The initial temperature is the ambient temperature of 295 K. At around 200 s, the temperature of point 1 increases when the laser starts to scan at that location. 
However, there are some discrepancies between the simulated and measured data during this initial increase. The measured temperature increases more slowly than the simulation and notably does not exceed the solidus temperature of the material (1533 K), which is incorrect. Additionally, in the IR images in Fig.\ref{fig:meshtoolpath}d and e, the top surface, especially the melt pool right under the laser source, appears cooler than some of the previously built layers. This failure to measure the temperature of the molten material is attributed to the difference in emissivity between the liquid and solid phases of the material \cite{yang2017thermal}. The emissivity of liquid metal alloys is much lower than that of solid metal alloys, leading to erroneously low-temperature measurements in the melt pool. Near and above the solidus temperature, therefore, the simulated temperatures are considered more reliable than experimental captured data. 

After the simulated temperature decreases below the solidus temperature, the simulated and experimental curves match well. The rapid oscillation of each temperature curve is caused by the multiple passes of the laser as additional material is added; both the mean and the amplitude of the oscillations decreases as the wall grows in height and the amount of material between the point and the laser spot increases. The laser source is turned off around 615 s in Fig.~\ref{fig4.1:TemperatureHistory_all}a, after which both the simulated and experimental curves decrease more quickly and monotonically. The cooling rate during this final period is mainly determined by free convection and radiation at the wall and substrate surfaces. Trends of thermal histories are similar in Fig.~\ref{fig4.1:TemperatureHistory_all}b-k.

\subsection{Microstructure analysis}
\begin{figure}
	\centering
	\includegraphics[width=6in]{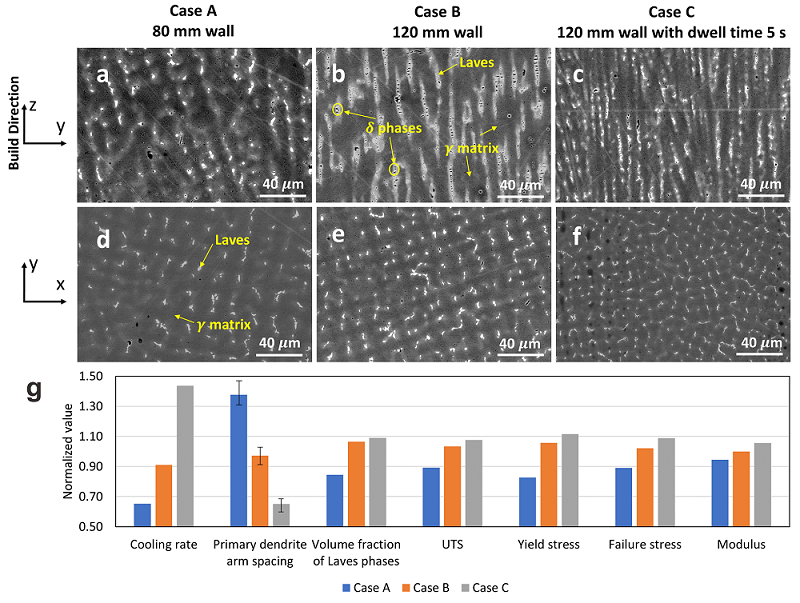}
	\caption{SEM characterization of the microstructure for parts with different process parameters. High magnification images \textbf{a}, \textbf{b},and \textbf{c} are the side view (normal to the scan direction) of samples from the three cases, while \textbf{d}, \textbf{e},and \textbf{f} show the top view (normal to the build direction). The locations of probed points are $(x,z)$ = (8.3,34.6) mm, (15,32.8) mm and (15.1,39.3) mm for Case A, B, and C respectively. The scale bar is 40 $\mathrm{\mu}$m. Some original SEM images (\textbf{a},\textbf{b},\textbf{c}) are repeated in \citet{bennett2021relating}. \textbf{g} Comparison of computed cooling rate, microstructure and mechanical properties of different thin walls. Values are tabulated in Supplementary Table 5, normalized by the mean across the three cases.}
	\label{fig5.1:SEM3612}
\end{figure}

The microstructure of DED thin walls is observed by the scanning electron microscope (SEM). The SEM-imaged microstructure for samples from the 80 mm thin wall, 120 mm thin wall, and 120 mm thin wall with dwell time 5 s (Cases A, B, and C) are shown in Fig.~\ref{fig5.1:SEM3612}a-f. The locations for the probed points in the three different thin wall parts are $(x,z) = $ (8.3, 34.6) mm, (15, 32.8) mm and (15.1, 39.3) mm, respectively, where the $(x,z) = (0,0)$ point is taken as the lower left corner of thin walls in Fig.~\ref{fig:meshtoolpath}d and e. The simulated temperature is recorded at the center through the thickness of each sample location. Microstructure SEM samples are imaged from the side view (normal to the scan direction, i.e. the y-z plane) and top view (normal to the build direction, i.e. the x-y plane). From the SEM images, we can see the $\gamma$ matrix phase, the Laves phases, and a few $\delta$ phases (black dots inside of Laves phases). Laves phases are often observed to precipitate at grain boundaries. The $\delta $ ($\mathrm{Ni}_{3}\mathrm{Nb}$) phases usually precipitate along grain boundaries and within the intergranular matrix. Both Laves phases and $\delta$ phases are detrimental for mechanical properties \cite{sui2019influence}. Using the image processing software ImageJ \cite{rasband1997imagej}, we calculated the volume fraction of Laves phases and primary dendrite arm spacing for the top and side view of microstructures in the three thin walls shown in Supplementary Tables 5 and 6. 

\subsection{Process and location effects on the cooling rate, microstructure and mechanical properties}

We compare cooling rates, microstructure features (primary dendrite arm spacing and the volume fraction of Laves phases), and experimentally measured mechanical properties (UTS, yield stress, failure stress, and modulus) in Supplementary Table 5 for three cases. These values are plotted in Fig.~\ref{fig5.1:SEM3612}g, normalized by the mean value across the three cases. The primary dendrite arm spacing is measured by calculating the mean value of several primary dendrite arm spacings in the SEM images in Fig.~\ref{fig5.1:SEM3612}. The 80 mm wall (Case A) has the largest primary dendrite arm spacing, while the 120 mm wall with 5 s dwell time (Case C) has the smallest. Error bars based on the standard deviation across measurements are also shown in Fig.~\ref{fig5.1:SEM3612}g. From the figure, we can see that Case C, with the highest cooling rate, has the finest grains and relatively high strength (UTS, yield stress, failure stress, and modulus), while Case A, with the lowest cooling rate, has coarser grains and lower strength. We can conclude that increasing the time between successive laser scans, whether by increasing the wall size or the dwell time, increases the cooling rate and leads to finer microstructures and higher strength. 

We also investigate variations in microstructure and properties at different locations in the same wall. Top and side views of the microstructure are imaged at three different locations for Case B in Fig.~\ref{fig5.3:sem8}a-f. Specimen locations are shown in Supplementary Figure 1; point 1 is closer to the top of the wall, point 8 is in the middle, and point 12 is near the bottom. Primary dendrite arm spacing and Laves phases volume fraction are calculated from the SEM images in Fig.~\ref{fig5.3:sem8} via ImageJ and listed in Supplementary Table 6 along with the cooling rate, yield stress, UTS, failure stress, and modulus. These values, normalized by their means, are plotted in Fig.~\ref{fig5.3:sem8}g. 

From the results, we see no clear correlation between cooling rate, microstructure, and mechanical properties. The middle location has larger primary dendrite arm spacing than the top and bottom specimen. From bottom to top, the primary dendrite arm spacing gradually increases but then decreases. The trend is the same with the volume fraction of Laves phases and mechanical properties. But cooling rate decreases monotonically from top to bottom of the wall although the differences in cooling rate between the three locations are small. These results indicate that the cooling rate is not a good indicator to predict mechanical properties for different locations in one wall. Other thermal history features may be needed to develop correlations between process and mechanical properties. 

\begin{figure}
	\centering\includegraphics[width=6in]{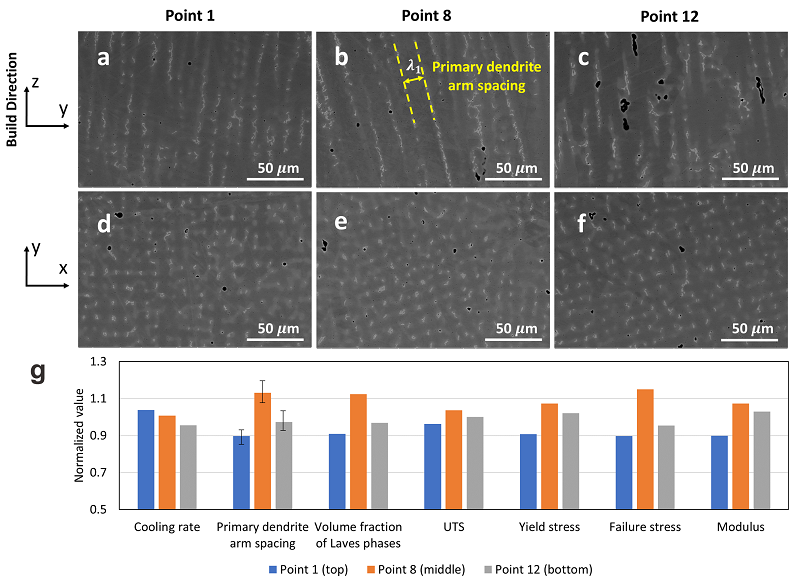}
	\caption{SEM microstructure images for different locations in Case B. High magnification images \textbf{a}, \textbf{b},and \textbf{c} are side views (normal to the scan direction) of the three locations noted in Supplementary Figure 1. Images \textbf{d}, \textbf{e},and \textbf{f} show top views (normal to the build direction) of the same locations. \textbf{g} Comparison of computed cooling rate, microstructure and mechanical properties of different locations in the wall for Case B. Values are tabulated in Supplementary Table 6, normalized by the mean across the three locations.}
	\label{fig5.3:sem8}
\end{figure}

\subsection{Prediction of primary dendrite arm spacing}
\begin{figure}
	\centering
	{
		\begin{minipage}[b]{3in}
			\centerline{\includegraphics[height=2.5in]{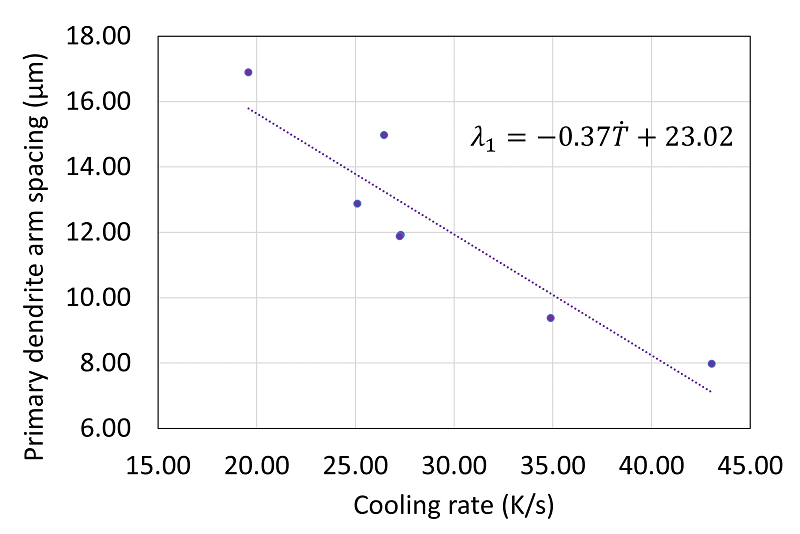}}	
			\centerline{(a)}
		\end{minipage}
	}
	\hfill
	{
		\begin{minipage}[b]{3in}
			\centerline{\includegraphics[height=2.5in]{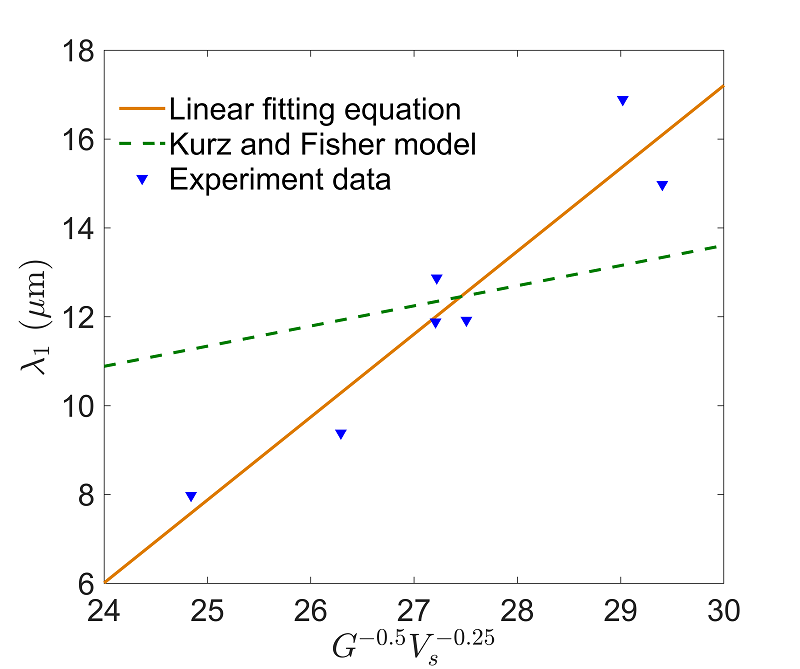}}	
			\centerline{(b)}
		\end{minipage}
	}
	\caption{Analysis of the primary dendrite arm spacing. (a) Primary dendrite arm spacing vs cooling rate. The dotted line is the linear fit, $\lambda_{1} = -0.37\dot{T}+23.02$, where $\lambda_{1}$ is the primary dendrite arm spacing  ($\mathrm{\mu m}$) and $\dot{T}$ is the cooling rate (K/s). (b) Comparison of measured primary dendrite arm spacing with analytical models. The solid line is the best fit to a model linear in $G^{-0.5} V_{s}^{-0.25}$ in Eq.~\ref{eqn:linearFit}, while the dashed line is the analytical model of Kurz and Fisher in Eq.~\ref{eqn:kurzFisherFit}.}
	\label{fig5.4:spacing_cooling}
\end{figure}

The relationship between measured primary dendrite arm spacing and simulated cooling rate for samples from different cases and locations is shown in Fig.~\ref{fig5.4:spacing_cooling}a. The primary dendrite arm spacing are measured from the 7 SEM microstructure images with locations of $(x,z)$ = (8.3,34.6) mm in Case A, $(x,z)$ =(15.0,32.8) mm in Case B, $(x,z)$ =(15.2,39.3) mm in Case B, $(x,z)$ =(73.2,29.8) mm in Case B, $(x,z)$ =(106.6,20.3) mm in Case B, $(x,z)$ =(14.1,39.1) mm in Case C, and $(x,z)$ =(15.1,39.3) mm in Case C. The simulated cooling rate are also calculated at the corresponding locations. 
The dotted line in Fig.~\ref{fig5.4:spacing_cooling}a is the linear fit to the data: 
\begin{equation}
	\lambda_{1}=-0.37\dot{T}+23.02
\end{equation} 
where $\lambda_{1}$ is the primary dendrite arm spacing ($\mathrm{\mu m}$) and $\dot{T}$ is the cooling rate (K/s). The $R^2$ error for the fit is 0.85. Results indicate a strong correlation between primary dendrite arm spacing and cooling rate. It shows that increasing cooling rate leads to smaller dendrite arm spacing, which is in agreement with previous findings \cite{li2015microstructure, popovich2017functionally}. 

According to solidification theory \cite{kurz1981dendrite}, primary dendrite arm spacing is closely related to thermal gradient and solidification rate. To understand the relationship between the primary dendrite arm spacing, thermal gradient and solidification rate, we compared our data with the analytical model of Kurz and Fisher \cite{kurz1981dendrite}:
\begin{equation}
	\lambda_{1}=A\left(\Gamma \Delta T_{0} D_{\ell} / k\right)^{0.25} G^{-0.5} V_{s}^{-0.25}
\end{equation}
where $\lambda_{1}$ is the primary dendrite arm spacing ($\mathrm{m}$), $A$ is a fitting coefficient, $\Gamma$ is the Gibbs-Thomson coefficient ($\Gamma = 3.65 \times 10^{-7}$ K$\cdot$m), $\Delta T_{0}$ is the undercooling ($\Delta T_{0}=T_l-T_s$, the unit is K), $D_{\ell}$ is the diffusion constant in the liquid ($\mathrm{m}^2/\mathrm{s}$), $k$ is the partition coefficient ($k = 0.48$ \cite{nie2014numerical}), $G$ is the thermal gradient (K/m), and $V_s$ is the solidification rate (m/s). The thermal gradient and solidification rate are calculated from the simulation data (as described in Methods) while the primary dendrite arm spacing is measured from experiments. We calculate the value of $A$ from the best fit to our experimental data using the Kurz and Fisher model. The final form of the Kurz and Fisher model based on our sample data is 
\begin{equation}
	\lambda_{1}= B\cdot G^{-0.5} V_{s}^{-0.25}
	\label{eqn:kurzFisherFit}
\end{equation}
where the coefficient $B=A\left(\Gamma \Delta T_{0} D_{\ell} / k\right)^{0.25}= 4.53 \times 10^{-7}$ $\mathrm{K}^{0.5} \cdot \mathrm{m}^{0.75}/\mathrm{s}^{0.25}$. 

We also fit the data to an expression that is linear in $G^{-0.5} V_{s}^{-0.25}$:
\begin{equation}
	\lambda_{1}=C_{1} \cdot G^{-0.5} V_{s}^{-0.25}+C_{2}
	\label{eqn:linearFit}
\end{equation}
where the coefficient $C_{1}=1.87 \times 10^{-6}$ $\mathrm{K}^{0.5} \cdot \mathrm{m}^{0.75}/\mathrm{s}^{0.25}$, and $C_{2}=-3.88\times 10^{-5}$ $\mathrm{m}$.

Fig.~\ref{fig5.4:spacing_cooling}b shows the comparison between the Kurz and Fisher model in Eq.~\ref{eqn:kurzFisherFit}, the linear fit in Eq.~\ref{eqn:linearFit}, and the experimental data. The major difference between the Kurz and Fisher model and the linear fit is that the linear model has an offset at the origin while the Kurz and Fisher model does not. The deviation of the Kurz and Fisher model might be a result of the rapid cooling rate and complex geometry of the dendrite tips instead of the simple assumptions of the model \cite{kirkaldy1995thin, liu1995thin, kurz1981dendrite, keller2017application}; it may also be caused by uncertainty in the measurement of the dendrite arm spacing. 

\subsection{Relationship between thermal history and mechanical properties}

\begin{figure}
	\centering
    {
		\begin{minipage}[b]{3in}
			\centerline{\includegraphics[width=3in]{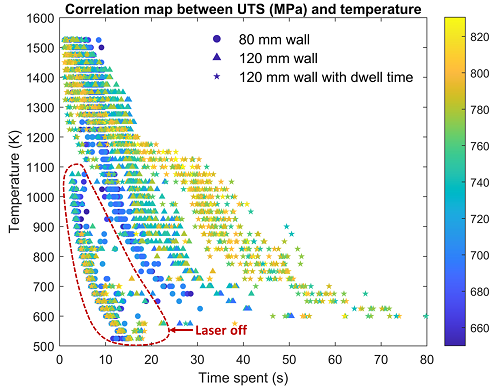}}	
			\centerline{(a)}
		\end{minipage}
	}
	\hfill
	{
		\begin{minipage}[b]{3in}
			\centerline{\includegraphics[width=3in]{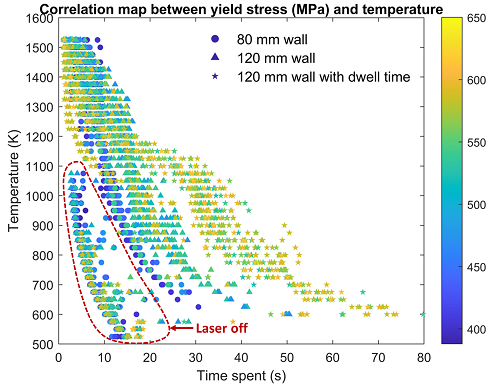}}	
			\centerline{(b)}
		\end{minipage}
	}
	\vfill
	{
		\begin{minipage}[b]{3in}
			\centerline{\includegraphics[width=3in]{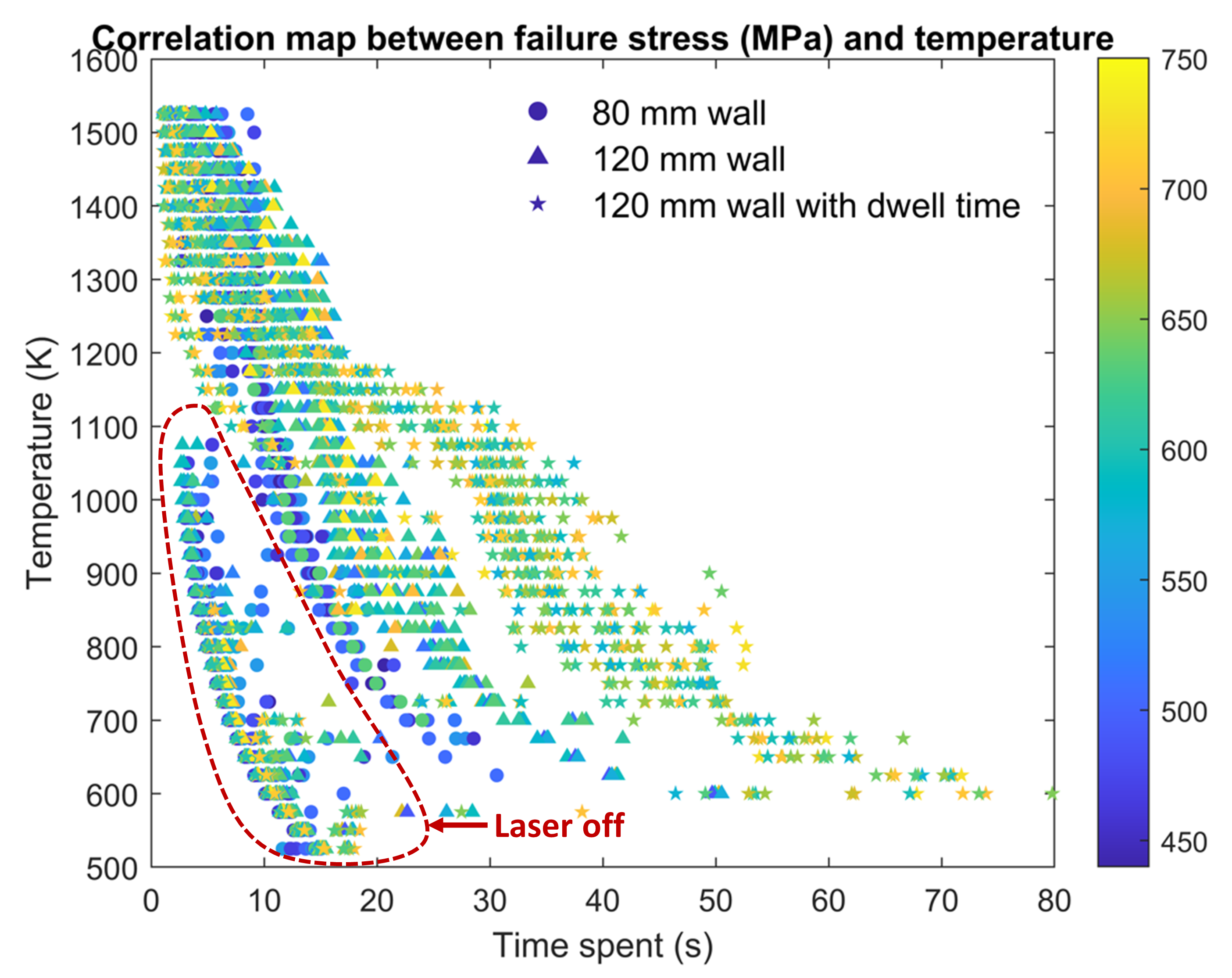}}	
			\centerline{(c)}
		\end{minipage}
	}
	\hfill
	{
		\begin{minipage}[b]{3in}
			\centerline{\includegraphics[width=3in]{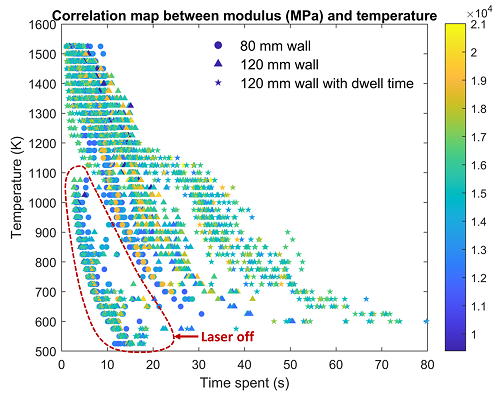}}	
			\centerline{(d)}
		\end{minipage}
	}
	\vfill
	{
		\begin{minipage}[b]{3in}
			\centerline{\includegraphics[width=3in]{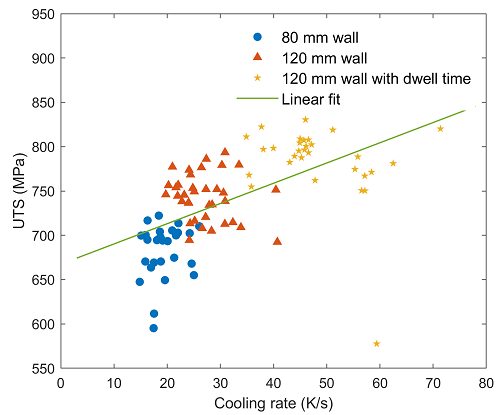}}	
			\centerline{(e)}
		\end{minipage}
	}	
	\caption{Correlation between temperature and mechanical properties for three different thin walls (80 mm wall, 120 mm wall, and 120 mm wall with 5 s dwell time). (a) UTS (b) yield stress (c) failure stress (d) modulus. The location of each marker shows how much time was spent within each 25-degree temperature band. Details of temperature post processing is described in Methods and Supplementary Figure 2. The color of each marker represents the value of the mechanical property for that sample. (e) Correlation between cooling rate and UTS for three different thin walls. The green line is the linear fit to all data. The $R^2$ error for the fit is 0.32.}
	\label{fig5.6:CorrMap}
\end{figure}

Fig.~\ref{fig5.6:CorrMap}a shows the correlation between thermal history and UTS for locations on all cases (Cases A, B, and C). Each point on the plot represents the cumulative time spent in a given temperature range for a particular simulated temperature history; the color of each point represents the experimentally measured UTS for the corresponding location, while the symbol shape (circle, triangle, or star) denotes the three different walls (Case A, B and C). Details of calculating cumulative time of thermal histories in each temperature band are described in Methods and Supplementary Figure 2. Case C (the 120 mm wall with 5 s dwell time) shows higher UTS than the walls without dwell time. It is conjectured that increasing dwell time during deposition leads to an increased cooling rate (refer to Fig.~\ref{fig5.1:SEM3612}g), resulting in finer microstructure and higher strength. 

For the temperature between 1200 and 1533 K in Fig.~\ref{fig5.6:CorrMap}a, less time is spent during each temperature band for Case C (with dwell time) than Case B (the same wall size without dwell time), which also indicates faster solidification in Case C. For temperatures below 1200 K, much more time is spent in each temperature range in Case C. Both shorter time spent during solidification and longer time spent at more moderate temperatures correlates with higher UTS. Likewise, Case B samples have higher UTS than Case A in Fig.~\ref{fig5.6:CorrMap}a, and comparatively more time is spent in each temperature range for Case B. This is caused by the longer wall length and scan path in Case B, giving longer times between successive laser passes.  

The zone enclosed in a red dashed line in Fig.~\ref{fig5.6:CorrMap}a represents the portion of the thermal histories after the laser is turned off. For the laser-off region, time spent is approximately the same for the three walls, indicating that differences in temperature history after laser shut-off do not play a role in the differences in mechanical properties in our tests. 

Similar correlation maps with point colors corresponding to yield stress, failure stress, and elastic modulus are shown in Fig.~\ref{fig5.6:CorrMap}b-d. Trends in yield and failure stress are similar to those of UTS. However, there is no apparent difference in modulus among the walls. This indicates that different microstructures caused by thermal history features may lead to variations in material strength, but minor effects on elastic modulus.

Fig.~\ref{fig5.6:CorrMap}e is the relationship between the cooling rate and UTS for three different thin walls. The green line in Fig.~\ref{fig5.6:CorrMap}e is the linear fit to all data. The $R^2$ error for the fit is 0.32. Results show that UTS increases as the cooling rate increases. Case C has the largest UTS with the largest cooling rate among the three cases. But the correlation between cooling rate and UTS for each case is weak, indicating that cooling rate by itself is not enough to predict mechanical properties. We need to consider more factors or features in the thermal histories besides cooling rate to predict mechanical properties.

\subsection{Prediction of mechanical properties using 1D CNN}

In the previous section, we discuss the trends in mechanical properties with respect to observable features of the temperature history, such as cooling rate and the reduced-order temperature feature descriptors (time spent during each temperature interval). However, the cooling rate alone is insufficient in predicting mechanical properties, and it is difficult to express a quantitative relationship between properties and the reduced-order temperature descriptors we highlight. Therefore, we aim to use machine learning to extract important features of the thermal histories and construct predictive models of the properties from a given thermal history.  

A convolutional neural network (CNN) can automatically extract essential features and can learn high-level features from spatial and temporal data \cite{herriott2020predicting, guo2016convolutional, kim2020prediction, fonda2019deep}. Recent work shows that one-dimensional CNN can be used to analyze time series or sequence data effectively \cite{abdeljaber20181, abdeljaber2017real, kiranyaz2015real, kiranyaz20191d,avci2018wireless, yang2019fault}. This work uses 1D CNN to extract features from thermal histories and predict mechanical properties, such as UTS, of the sample points. The data preparation, hyperparameter search, neural network architecture, and results are discussed in Methods.

\begin{figure}
	\centering
	\includegraphics[width=6in]{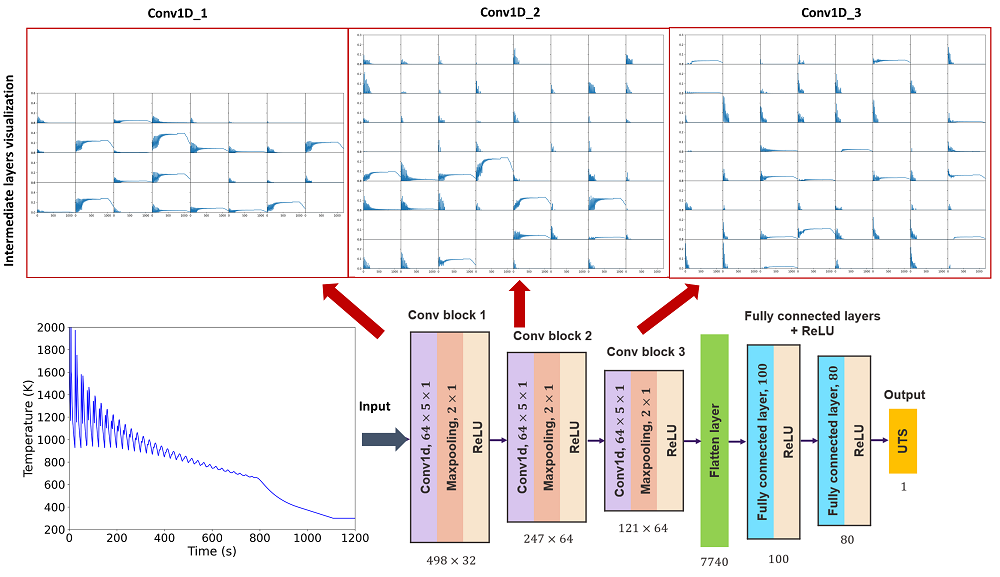}
	\caption{The CNN structure with intermediate layer visualization. The CNN architecture consists of three convolutional layers with max pooling layer and the ReLU activation, and three fully connected layers with the ReLU activation. The top three subplots are the visualization of output of the first, second, and third convolutional layers. The inputs of 1D CNN are the pre-processed thermal histories. The output is mechanical properties (UTS).}
	\label{fig5.7:cnn_interlayer_structure}
\end{figure}

\begin{figure}
	\centering
	{
		\begin{minipage}[b]{3.1in}
			\centerline{\includegraphics[width=2.8in]{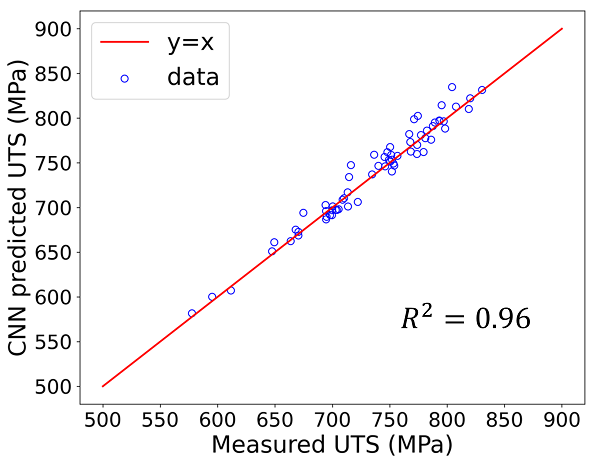}}	
			\centerline{(a)}
		\end{minipage}
	}
	\hfill
	{
		\begin{minipage}[b]{3.1in}
			\centerline{\includegraphics[width=2.8in]{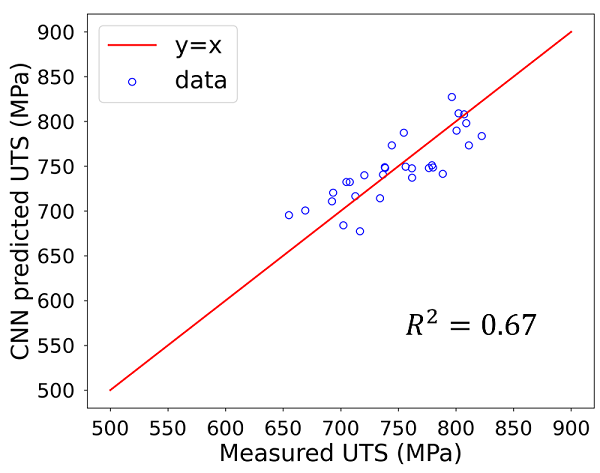}}	
			\centerline{(b)}
		\end{minipage}
	}
	\caption{Measured UTS versus predicted UTS. (a) Training data. (b) Test data. }
	\label{fig5.7:CNNresults}
\end{figure}

 Fig.~\ref{fig5.7:cnn_interlayer_structure} is the convolutional neural network structure with the intermediate convolutional layer visualization. The input of the trained CNN is the temperature history for each probed location in the deposited thin-wall. The output of the network is UTS for corresponding locations. Fig.~\ref{fig5.7:CNNresults} shows the comparison of CNN-predicted UTS with the actual measured UTS. Fig.~\ref{fig5.7:CNNresults}a is the training data while Fig.~\ref{fig5.7:CNNresults}b is the test data. The $R^{2}$ scores for the training and testing are 0.96 and 0.65, as shown in the plot. The results show that the proposed CNN structure can accurately predict the UTS for samples in these thin-wall builds based on thermal history.

To understand which features of the thermal processes have a dominant effect on mechanical properties, we output the intermediate convolutional layers of the trained network. We visualize the output of the first, second, and third convolutional layer (Conv1D\textunderscore1, Conv1D\textunderscore2, Conv1D\textunderscore3) for all feature filters; these are displayed above the CNN model in Fig.~\ref{fig5.7:cnn_interlayer_structure}. The channel outputs of each convolution layer inform how the convolutional layers extract the key factors from the temperature history and transfer useful information to formulate the mechanical properties. There are 32, 64 and 64 channels in the Conv1D\textunderscore1, Conv1D\textunderscore2, and Conv1D\textunderscore3 layers, respectively. 

\begin{figure}
	\centering
	{
		\begin{minipage}[b]{3in}
			\centerline{\includegraphics[width=4.5in]{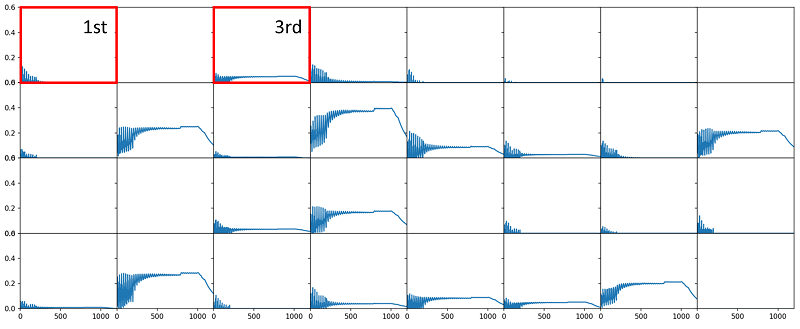}}	
			\centerline{(a)}
		\end{minipage}
	}
	\vfill
	{
		\begin{minipage}[b]{3in}
			\centerline{\includegraphics[width=3in]{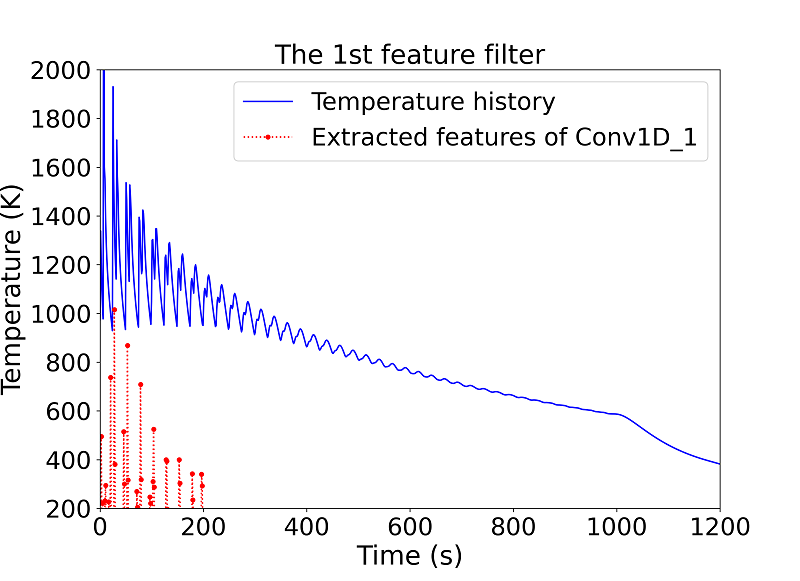}}	
			\centerline{(b)}
		\end{minipage}
	}
	\hfill
	{
		\begin{minipage}[b]{3in}
			\centerline{\includegraphics[width=3in]{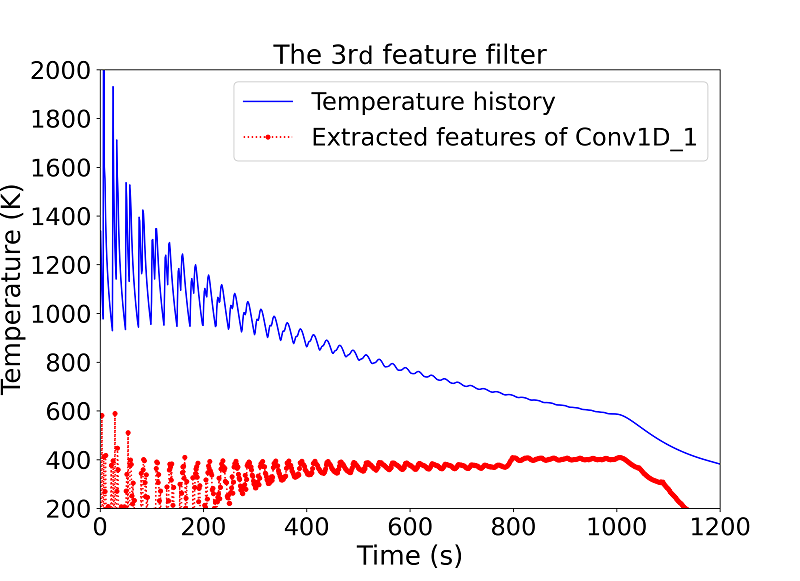}}	
			\centerline{(c)}
		\end{minipage}
	}
	\caption{The output of the first convolutional layer for all channels. Each subplot in (a) represents the output of the first convolutional layer for each channel/filter. The total number of channels is 32. (b) Comparison of temperature history with extracted features from the first convolutional layer with the first feature filter; (c) Comparison of temperature history with extracted features from the first convolutional layer with the third feature filter. The blue line represents temperature history of one case. The red dots are the extracted features from one feature filter in the first convolutional layer. The original extracted feature values are multiplied by 8000 for convenient comparison with temperature history in the same plot.}
	\label{fig5.7:cnn1_output}
\end{figure}

Fig.~\ref{fig5.7:cnn1_output} shows the output from the first convolutional layer. Fig.~\ref{fig5.7:cnn1_output}a is the output of the first convolutional layer for all channels. There are 32 feature filters for the Conv1D\textunderscore1. The non-zero output plots display two main patterns. One pattern, displayed for example by the first feature filter, has the most values in the top few time steps. The other pattern has values from 0 s to around 1100 s, as seen for example in the third feature filter. Some other outputs in Fig.~\ref{fig5.7:cnn1_output}a are not activated and close to zero, which means that these feature filters do not capture trends from the input and do not transfer any information to the next layer in the network. A comparison of the extracted features output for the first and third feature filters with temperature history are shown in Fig.~\ref{fig5.7:cnn1_output}b and c, respectively. The blue line represents the temperature history. The red dot line represents the extracted output of the first convolutional layer for the first and third feature filters. The original extracted feature values are multiplied by 8000 for convenient comparison with temperature history in the same plot. The results indicate not only the temperature during the first few cycles (the solidification range) influences on mechanical properties, but also the entire temperature history from the start of the laser scan to the end of the process (around 1100 s) plays an important role. The cooling rate calculated from the solidification period cannot represent all thermal features on the effect of mechanical properties. Both the solidification temperature and whole repeated thermal cycle during laser scan affect the formation of microstructure and final mechanical properties, such as UTS.

\begin{figure}
	\centering
	{
		\begin{minipage}[b]{3in}
			\centerline{\includegraphics[width=3in]{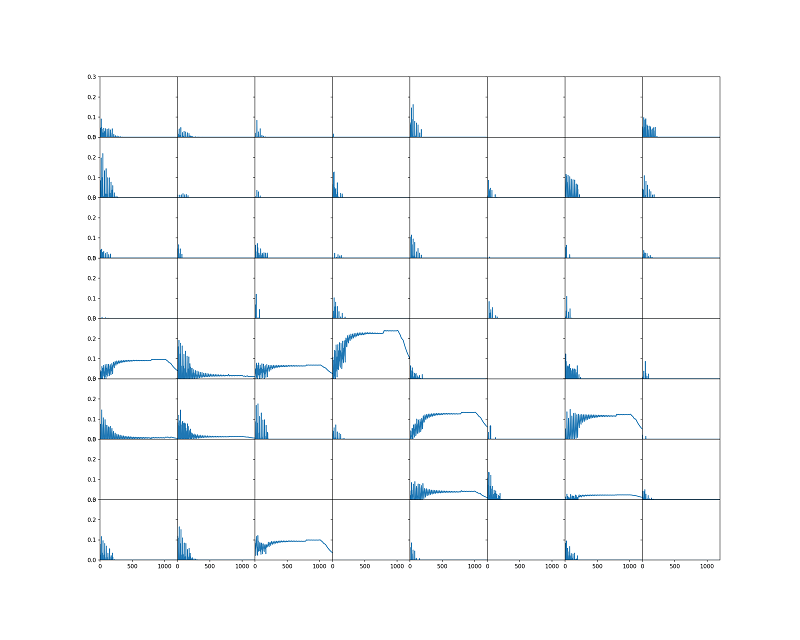}}	
			\centerline{(a)}
		\end{minipage}
	}
	\hfill
	{
		\begin{minipage}[b]{3in}
			\centerline{\includegraphics[width=3in]{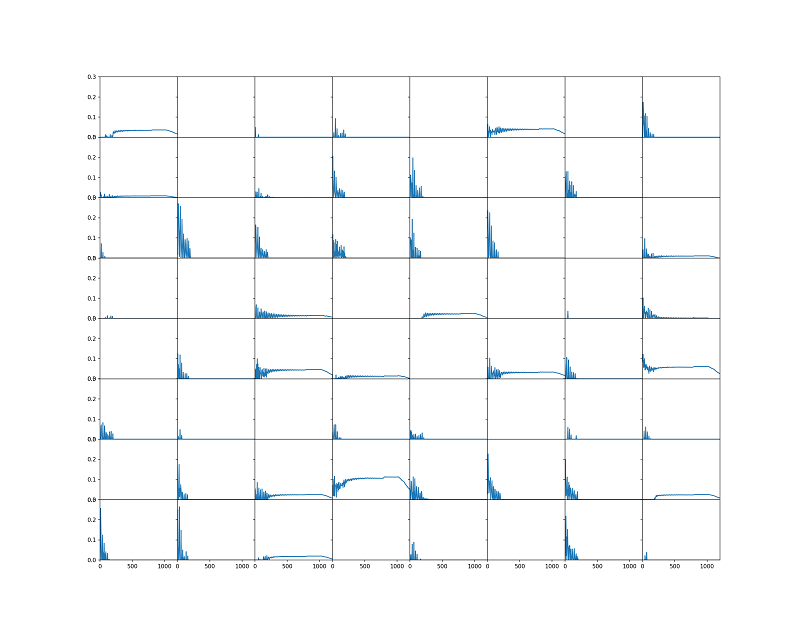}}	
			\centerline{(b)}
		\end{minipage}
	}	
	\vfill
	{
		\begin{minipage}[b]{3in}
			\centerline{\includegraphics[width=3in]{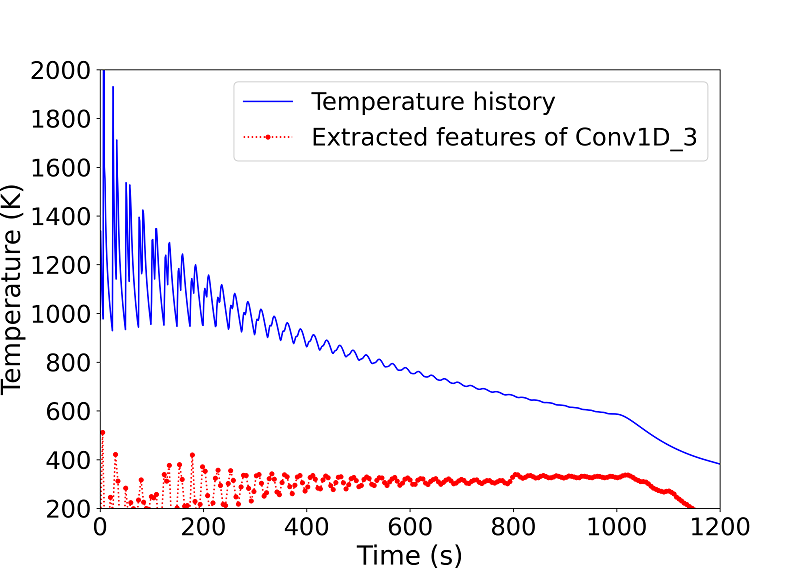}}	
			\centerline{(c)}
		\end{minipage}
	}
	\hfill
	{
		\begin{minipage}[b]{3in}
			\centerline{\includegraphics[width=3in]{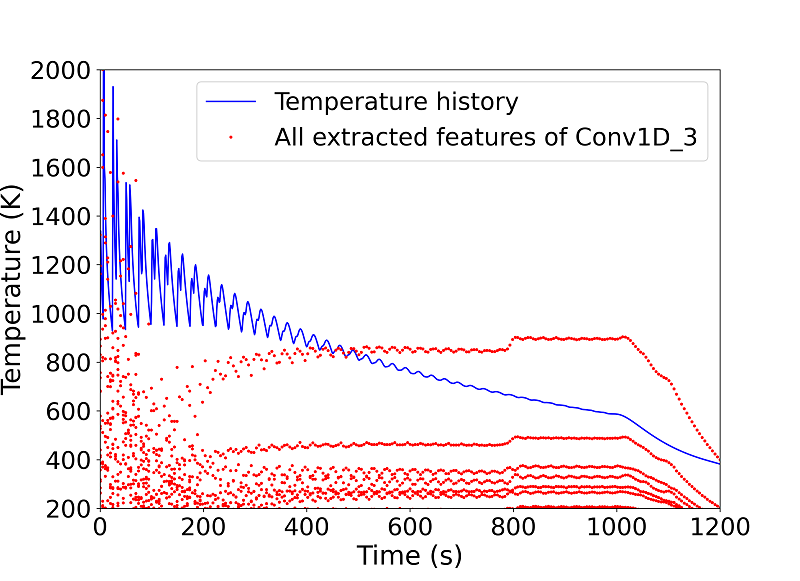}}	
			\centerline{(d)}
		\end{minipage}
	}
	\caption{The output of the second and third convolutional layers. (a) Second convolutional layer. (b) Third convolutional layer. Each subplot in (a) and (b) represents the output of the layer for a single one of the 64 channels/filters. (c,d) Comparison of temperature history with extracted features from the third convolutional layer for (c) a single feature filter and (d) all feature filters. The blue lines are the temperature history of a representative case. The red dots are the extracted features from feature filters in the third convolutional layer. The original extracted feature values are multiplied by 8000 for convenient comparison with temperature history in the same plot.}
	\label{fig5.7:cnn23_output}
\end{figure}

We also show the output of the second and third convolutional layers in Fig.~\ref{fig5.7:cnn23_output}a and Fig.~\ref{fig5.7:cnn23_output}b. There are 64 feature filters for Conv1D\textunderscore2 and Conv1D\textunderscore3. Comparing the outputs from the Conv1D\textunderscore1, Conv1D\textunderscore2, and Conv1D\textunderscore3, the first convolutional layer captures most of the information from the input, although some feature filters are not activated. For deeper layers, the output visualization becomes more abstract and less interpretable. The Conv1D\textunderscore3 contains less information in the output visualization, with mostly zero output for a number of feature filters. 

The comparison of the extracted features of the Conv1D\textunderscore3 for one chosen feature filter with temperature history is presented in Fig.~\ref{fig5.7:cnn23_output}c, and all feature filters are shown in Fig.~\ref{fig5.7:cnn23_output}d. In Fig.~\ref{fig5.7:cnn23_output}c, the extracted features capture the temperature information from 0 s to around 1150 s. Looking at extracted features from all feature filters in Conv1D\textunderscore3 in Fig.~\ref{fig5.7:cnn23_output}d, the output captures temperature information from the start of the laser scan to the end of laser fusion, similar to the trend of the outputs from Conv1D\textunderscore1 and Conv1D\textunderscore2. 

\section{DISCUSSION}
\label{S:6}
In this paper, we developed a thermal model using a finite element method with spatially varied surface convection, which can accurately predict the temperature field of DED builds for Inconel 718. The thermal model was validated using in situ experiments. Microstructure characterizations and tensile tests were conducted at material samples throughout the built parts to further understand the process-structure-property relationship. Furthermore, a data-driven model based on a 1D convolutional neural network was used to efficiently predict mechanical properties from thermal histories. 

The results demonstrate that the thermal model can accurately predict the temperature field for a DED process. The thermal history features can be related to simulated cooling rate, dendrite arm spacing, and UTS for the DED thin-wall parts. Increasing dwell time and length of deposition leads to a finer microstructure and associated changes in mechanical properties, such as higher strength. From the correlation of cooling rate with UTS, we find that the cooling rate calculated based on the solidification period alone is not enough to predict mechanical properties.

An advanced machine learning technique (1D convolutional neural network) offers an efficient way to automatically extract features from input data without a priori feature selection and predict mechanical properties from a time series. With a trained CNN, simulated thermal histories rather than cooling rates can give good predictions of mechanical properties. From visualization of intermediate convolutional layers, it is observed that not only the solidification range but also the entire thermal history during the laser scan has effects on final mechanical properties. This might indicate that some precipitates, such as Laves phases, $\delta$ phase ($\mathrm{Ni}_{3}\mathrm{Nb}$), $\gamma^{\prime}$ phase $\mathrm{Ni}_{3}(\mathrm{Al},\mathrm{Ti})$ and $\gamma^{\prime \prime}$ phase ($\mathrm{Ni}_{3}\mathrm{Nb}$), form during the intermediate temperature range. $\gamma^{\prime \prime}$ phase is the major strengthening precipitation in Inconel 718 alloy while $\gamma^{\prime}$ phase is an assistant strengthening precipitate \cite{sui2019influence}. The $\delta$ phase and Laves phases are detrimental to material properties (strength, fatigue life, and ductility) \cite{anderson2017delta}. 

Different precipitates can form in two stages: non-equilibrium solidification and solid-state phase transformation after solidification. In non-equilibrium solidification, the $\gamma$ matrix begins to solidify when the temperature is below the liquidus temperature during the first several thermal cycles. The segregation of Inconel 718 component elements changes the driving force of phase formation, which leads to the formation of Laves phases in the interdendritic region \cite{kumara2020toward}. It has been reported that Laves formation is related to slow solidification at a low cooling rate \cite{antonsson2005effect}. However, it is not sufficient to use cooling rate alone as the criterion of Laves formation because Laves phases are also observed at high cooling rates in AM processes \cite{ghosh2018formation, ghosh2018single}. Thermal gradients, undercooling and solid-liquid interface velocity should also be considered. As the temperature continues to drop after non-equilibrium solidification, solid-state phase transformation occurs. Precipitates such as $\delta$, $\gamma^{\prime}$ and $\gamma^{\prime \prime}$ phases may form in this stage. If time is sufficient between deposition of layers (e.g., for long dwell time), temperature decreases quickly, and the time that material spends in the precipitation formation temperature range (875 K to 1275 K) from the TTT (Temperature-Time-Transformation) diagram of Inconel 718 \cite{oradei1991current,sui2019influence} is too short to form $\delta$, $\gamma^{\prime}$ and $\gamma^{\prime \prime}$ \cite{kumara2020toward}. If the dwell time is not long enough to cool the material quickly, time spent in the precipitation temperature range may be long enough to form $\delta$, $\gamma^{\prime}$ and $\gamma^{\prime \prime}$ in the microstructure. Note that the time required to form precipitates has been observed to be shorter for AM than for wrought Inconel 625 due to increased interdendritic segregation in AM material \cite{lindwall2019simulation,stoudt2018influence,zhang2018effect} . In Inconel 718, the $\gamma^{\prime}$ phases and $\gamma^{\prime \prime}$ precipitation temperature range is from 875 K to 1175 K, while that for $\delta$ phases is from 1225 K to 1275 K \cite{anderson2017delta,kuo2017effect,li2002isolation}. Both temperature ranges for precipitation are within the intermediate temperature range identified from the visualization of our CNN feature filters. Therefore, CNN predictions may give more insights for future models to predict the formation and evolution of precipitates and correlate with mechanical properties.

One limitation of the CNN model is that its transferability to other geometries of parts is unknown. Machine learning models generally cannot be extrapolated outside the range of parameters for which they were trained. Therefore, additional experimental data for other geometries and different build parameters would help build a model that could be used to predict the strength of Inconel 718 based on thermal history for general DED build processes. However, the main insight gained through the CNN in this work, namely that the temperature history through intermediate temperature ranges (and not just solidification) has a measurable effect on material properties, is valuable information in the design and improvement of DED processes.

\section{METHODS}
\subsection{DED experiment setup and material}
Single-track thin walls of Inconel 718 were deposited via DED onto stainless steel 304 substrates using a hybrid additive and subtractive DMG MORI Lasertec 65 machine equipped with an in situ infrared melt pool monitor. The walls were produced using three different process conditions to vary the solidification and cooling rates between different cases: 80 mm long walls (Case A), 120 mm long walls (Case B), and 120 mm long walls with a 5 s inter-layer dwell time (Case C). Three repeated experiments are conducted for each case. Each wall is built with 120 layers. The laser power is 1800 W, and the laser scan speed is 16.7 mm/s for all builds. The design of the laser tool path in the experiment is shown in Fig.~\ref{fig:meshtoolpath}b. The powder mass flow rate is 0.3 $\mathrm{g}/\mathrm{s}$ with a powder focus radius of 3 mm. For Case C, after the deposition of each layer, a 5 s dwell time (a pause in the build) is applied before beginning the next layer, allowing for additional cooling.

\subsection{IR thermal measurement}
In the deposition process, a FLIR A655sc digital IR camera was used to capture thermal images of the thin wall during deposition. The IR camera recorded the infrared radiation during the process; temperature measurements were then obtained from the raw IR radiation data by calibrating the emissivity at the measured points. The resolution of the camera is $640 \times 480$ pixels, and the spectral range is from 7.5 to 14.0 $\mathrm{\mu}$m with accuracy of $\pm 2\; ^{\mathrm{o}}\mathrm{C}$. The recorded temperature range of the IR camera is from 300 $^{\mathrm{o}}\mathrm{C}$ to 2000 $^{\mathrm{o}}\mathrm{C}$. Additional details of the AM experiment and data processing from the IR camera are given in \cite{bennett2021relating,glerum2021mechanical,xie2021mechanistic}.

\subsection{Mechanical properties measurement}
Coupon specimens were cut via wire EDM for each wall, and measurements were performed for modulus, UTS, yield stress, yield strain, and failure stress. Vertically-oriented miniaturized ASTM E8 tensile specimens were manufactured from 9 different locations on each 80 mm wall (Case A) and 12 different locations on each 120 mm wall (Case B and Case C).  The temperature monitored by the IR camera and the locations of coupon specimens for tensile tests are shown in Fig.~\ref{fig:meshtoolpath} for reference; Fig.~\ref{fig:meshtoolpath}d shows locations of coupon specimens in Case A, while Fig.~\ref{fig:meshtoolpath}e shows the locations of specimens in Cases B and C. The tensile specimens were tested under displacement control to complete failure on a Sintech 20 G tensile test machine. 


\subsection{Microstructure characterization}
Subsets of the tensile specimens from selected wall locations were cut, mounted, and polished to a 0.02 $\mathrm{\mu}$m finish with non-crystallizing colloidal silica on a vibratory polisher. Microstructures at two faces of samples were observed, normal to the build direction and the scan direction, with a FEI Quanta 650 scanning electron microscope (SEM) with secondary electron (SE), backscatter electron (BSE), and energy-dispersive x-ray spectroscopy (EDS) detectors. Binarization and analysis of the SEM images were performed by ImageJ Software \cite{rasband1997imagej}.

\subsection{Thermal model}
The thin-wall builds are simulated using a finite element model. The thermal history during the DED process is calculated by solving the heat conduction equation in the following form:
\begin{equation}
\label{eq:heatconduction}
\rho C_{p} \frac{\partial T}{\partial t}(\mathbf{x}, t) = -\nabla \cdot \mathbf{q}(\mathbf{x}, t)
\end{equation} 
where $\rho$ is the material density, $C_p$ is the specific heat capacity, $T$ is temperature, $t$ is time, $\mathbf{q}$ is heat flux, and $\mathbf{x}$ is the spatial coordinate. 

The heat flux vector in Eq. (\ref{eq:heatconduction}) is assumed to be given by Fourier's Law:
\begin{equation}
    \mathbf{q}=-k\nabla T
\end{equation}
where $k$ is the thermal conductivity of the material, modeled as isotropic (so that $k$ is a scalar). The thermal properties of Inconel 718, namely the specific heat capacity and thermal conductivity, are taken as temperature-dependent with values listed in Supplementary Table 7. Properties at temperatures between those in the table are calculated by linear interpolation.

\subsection{Initial and boundary conditions}
The initial condition for the thin wall deposition and substrate is a constant temperature: 
\begin{equation}
	T(\mathbf{x},t_0)=T_{\infty}
\end{equation}
where $T_{\infty}$ is the ambient air temperature. The boundary condition describes the heat flux at the surface of the domain, which includes the heat loss due to radiation, convection, evaporation, and heat source from laser. The general form of the heat flux boundary condition is
\begin{equation}
    \mathbf{q} \cdot \mathbf{n}=q_{\mathrm{rad}}+q_{\mathrm{conv}}+q_{\mathrm{evap}}+q_{\mathrm{laser}} \qquad \mathrm{on} \; \Gamma_{q}
\end{equation}
where $\Gamma_{q}$ is the boundary surface with normal vector $\mathbf{n}$, $q_{\mathrm{rad}}$ represents the radiative heat flux applied on all exposed surfaces of the thin wall and substrates, $q_{\mathrm{conv}}$ is the heat flux due to convection on all surfaces (free convection with constant coefficient on the bottom of the substrate, and forced convection with spatially varying coefficient on the surface of the walls and other surfaces of substrate), $q_{\mathrm{evap}}$ is the heat flux due to evaporation, and $q_{\mathrm{laser}}$ is the laser heat flux that is applied on the exposed surfaces of the thin wall and the top surface of the substrate. Note that the exposed surfaces change throughout the build process as material is added. During the simulated build, elements representing newly added material are activated on a schedule determined by the laser path. Only active elements are considered in the computation, and the above boundary condition is applied at the outer boundary of the active domain at each time step.  

The heat flux boundary condition due to radiation exchange with the ambient surroundings is applied on all exposed surfaces of the thin wall and substrate: 
\begin{equation}
\label{eq:radiation}
q_{\mathrm{rad}}=\varepsilon \sigma\left(T^{4}-T_{\infty}^{4}\right)
\end{equation}
where $\varepsilon$ is the material emissivity and $\sigma$ is the Stefan-Boltzmann constant.

The heat flux due to convection is 
\begin{equation}
	q_{\mathrm{conv}}=h(x,y,z)\left(T-T_{\infty}\right)
\end{equation}
where $h$ is the heat convection coefficient, $x$, $y$, and $z$ are the coordinates of the point of interest. At the bottom surface of the substrate, constant free convection is applied with convection coefficient of 10 $\mathrm{W}/(\mathrm{m}^{2}\cdot \mathrm{K})$ to approximate free convection in air \cite{he2009temperature, pratt2008residual, wang2008residual, zheng2008thermal1, zheng2008thermal2}. At the exposed surfaces of thin wall and the top surfaces of the substrate, we use the same form of spatially varied heat convection coefficient model as \citet{heigel2015thermo}. The spatially varying convection model presented by \citet{heigel2015thermo} accounts for the effects of forced convection caused by the shield and carrier gas flows in the DED process. The heat convection coefficient on the wall surfaces is in the form: 
\begin{equation}
	h_{\mathrm{wall}}(x,y,z) = \left\{\begin{array}{ll}
		(-2.0 \Delta z+15) \mathrm{e}^{-(0.107 \sqrt{(x-x_b)^2+(y-y_b)^2+(z-z_b)^2})^{2.7}}+10 & \Delta z \leq 7.0 \;\mathrm{mm} \\
		10 & \Delta z>7.0 \;\mathrm{mm}
	\end{array}\right.
\end{equation}
where $\Delta{z}$ is the vertical distance from the top surface of the wall to the point of interest, $x$, $y$, and $z$ are the coordinates of the point of interest, and $x_b$, $y_b$, and $z_b$ are the coordinates of the center of the laser beam. 

Since convection on the vertical surfaces of the thin wall is expected to be different from that on the horizontal substrate surfaces, a slightly different model is applied to the substrate \cite{heigel2015thermo}. We also use the same form of the equation as \citet{heigel2015thermo} for the convection on the top surface of the substrate stated below:
\begin{equation}
	h_{\mathrm{substrate}}(x,y,z)=\left\{\begin{array}{ll}
		1.9(-2.0 \Delta z+15) \mathrm{e}^{ -(0.031 \sqrt{(x-x_b)^2+(y-y_b)^2+(z-z_b)^2})^{1.4}}+10 & \Delta z \leq 7.0\; \mathrm{mm} \\
		10 & \Delta z>7.0 \;\mathrm{mm}
	\end{array}\right.
\end{equation}
The parameters in both of these expressions for convection coefficient are calibrated using the DED experiment data. 

The heat loss due to evaporative cooling when material reaches the evaporation temperature is
\begin{equation}
	\label{eq:evap}
	q_{\mathrm{evap}}=-m_{\mathrm{evap}} L_{\mathrm{v}}
\end{equation}
where $m_{\mathrm{evap}}$ is the mass evaporation flux and $L_{\mathrm{v}}$ is the latent heat of vaporization. 
The mass evaporation flux term in Eq. (\ref{eq:evap}) follows the model proposed by Anisimov \cite{anisimov1995instabilities}:
\begin{equation}
	m_{\mathrm{evap}}=P_{\mathrm{sat}}(T)\left(\frac{m_{\mathrm{mol}}}{2 \pi R_{\mathrm{gas}} T}\right)
\end{equation}
where $P_{\mathrm{sat}}$ is the saturation pressure at temperature $T$, $R_{\mathrm{gas}}$ is the gas constant, and $m_{\mathrm{mol}}$ is the molar mass of the evaporating species. Specifically, the saturation pressure $P_{\mathrm{sat}}$  can be computed by solving the Clausius-Clapeyron equation  \cite{anisimov1995instabilities}: 
 \begin{equation}
 	P_{\mathrm{sat}}(T)=P_{\mathrm{a}} \exp \left(\frac{-L_{v} m_{\mathrm{mol}}}{R_{\mathrm{gas}}}\left(\frac{1}{T}-\frac{1}{T_{\mathrm{b}}}\right)\right)
 \end{equation}
where $T_{b}$ is the boiling point or the evaporation temperature at the ambient pressure $P_{\mathrm{a}}$. 

The laser heat flux boundary condition is applied on the exposed surfaces of the thin wall and the top surface of the substrate. The distribution of laser heat source is assumed to follow a Gaussian relationship in the following form:
\begin{equation}q_{\mathrm{laser}}=\frac{2 \eta P}{\pi r_{b}^{2}} \exp \left(\frac{-2 \left((x-x_{b})^{2}+(y-y_{b})^{2}+(z-z_{b})^{2}\right)}{r_{b}^{2}}\right)\end{equation}
where $\eta$ is the laser absorption efficiency, $P$ is the laser powder, and $r_b$ is the beam radius.

\subsection{Simulation cases and integration method}
The geometry of the computational domain and meshes are shown in Fig.~\ref{fig:meshtoolpath}c. The dimensions of the thin walls are listed in Supplementary Table 1. The dimension of the substrate in each case is $200\; \mathrm{mm} \times 75\; \mathrm{mm} \times75$ mm in the $x$, $y$, and $z$ direction. For each case, there are 120 deposition layers with a layer thickness of 0.5 mm. The laser power is 1800 W with the laser scan speed of 16.7 mm/s.  For Case A, the mesh contains 174805 elements with 222304 nodes; for Cases B and C, 238613 elements with 300036 nodes. For each layer, the mesh is discretized to contain one element in the building direction and six elements through the width of the wall. One element per layer for thin wall build mesh is found to be enough for computational accuracy since the depth of the melt pool is deeper than a single layer in the DED process. The time step size is 0.12 s for all cases. An element activation method is used to accommodate material addition during the DED process. The entire mesh is stored from the beginning of the simulation, but elements representing as-yet unbuilt material are inactive and not included in the finite element matrix assembly. Elements are activated according to the laser path and build schedule; as new elements are activated, the exposed surface on which boundary conditions are applied is also modified. More details of the computational approach can be found in \cite{wolff2017framework}. An 8-point Gaussian quadrature scheme (2 points in each direction) is used to evaluate spatial integrals in the finite element weak form. A backward Euler method is used to discretize the temporal term in Eq. \ref{eq:heatconduction}. The Newton-Raphson method is used to solve the resulting equation set, which is nonlinear because of the radiation terms in Eq. \ref{eq:radiation}.

\subsection{Thermal history post-processing and cooling/solidification rates}

During the AM process, the variation of mechanical properties is significant due to different thermal histories. To study the correlation between cooling time and mechanical properties (e.g., Fig.~\ref{fig5.6:CorrMap}a-d), we extract a reduced set of data features from each thermal history by dividing each thermal history into different temperature ranges with 25 K intervals and then calculating the cumulative time spent in each temperature range after solidification (see Supplementary Figure 2). Since we mainly focus on the thermal effects on mechanical properties during the material solidification period, we only consider the temperature history after the temperature decreases below the solidus temperature for the final time (the thermal history in the white area in Supplementary Figure 2).

It is expected that the cooling rate affects the formation of microstructure and final mechanical properties \cite{zhang2009effect, wolff2019experimentally, wei2019three}. In this work, we calculate the cooling rate as the average rate of change of temperature between the solidus temperature (1533 K) and an arbitrarily chosen temperature of 1450 K for all probed points. An example of the cooling rate calculation is shown in Supplementary Figure 3.

Since the primary dendrite arm spacing is related to the solidification rate and thermal gradient, we calculated the solidification rate from the simulated temperature fields as shown in Supplementary Figure 4. The solidification rate is in the direction of the normal vector to the surface of the liquid-solid interface, which is divided by the red and orange colors. The solidification rate $V_s$ can be calculated by $V_{s}=V\cos\theta$, where $V$ is laser scan speed, and $\theta$ is the angle between laser scan direction and the direction of the normal vector to the liquid-solid interface. Since each point in the thin wall undergoes repeated thermal cycles, we only focus on process-structure-properties during the solidification period and calculate the solidification rate at the last solidus temperature for the location. 

\subsection{Convolutional neural network}
We use the 1D CNN to extract features from thermal histories and predict UTS. The input of the CNN model is the thermal history at a given location in the form of a series of temperature values at regularly spaced points in time. The output is UTS for corresponding locations. Since the IR-measured temperature does not capture the temperature through the important initial solidification range accurately, as shown in Fig.~\ref{fig4.1:TemperatureHistory_all}, we use simulated thermal histories instead of experimental data as input in the model. To prepare the thermal history curve at each sample point, we define the initial time $t=0$ as the time when the laser spot first reaches that location for each temperature history, i.e., when the material is first added at that location. Note that the CNN requires each set of input data to have the same size. To meet this requirement, we "pad" the data to a uniform maximum time using a linear extrapolation down to the final constant ambient temperature, as shown in Supplementary Figure 5. To reduce the computational cost, we sample every 10th point from the original data so that points are separated by a time interval of $\Delta t=1.2$ s. We find that 1000 points provide enough resolution to define the temperature history curve accurately.

The CNN architecture used in this paper consists of three convolutional blocks followed by three fully connected layers, as shown in Fig.~\ref{fig5.7:cnn_interlayer_structure}. A $5 \times 1$ kernel size is chosen for each convolutional layer. Max-pooling layer is applied in each convolutional block, and a rectified linear unit (ReLU) activation is used for all convolutional and fully connected layers. The first two fully connected layers have 100 and 80 neurons, respectively. The input data size is 1000, which is the length of the series of temperature values in time. The output size is 1. Both the input and output data are min-max normalized before loading into the CNN training. The Adam optimizer is used in training to minimize the mean squared error (MSE). An early stopping method is applied to prevent overfitting. The patience of the early stopping is 100 epochs, which means that the training will stop if the validation loss does not improve after 100 epochs. The data set is split into 70\% for training and 30\% for testing, and the test data is used to measure validation loss. 

The MSE is defined in the following format:
\begin{equation}
\mathrm{MSE}=\frac{1}{n} \sum_{i=1}^{n}\left(y_{i}-\hat{y}_{i}\right)^{2}
\end{equation}

To evaluate the performance of the CNN structure, the $R^2$ score is calculated based on the difference between predicted and measured UTS:
\begin{equation}
R^{2}=1-\frac{\sum_{i=1}^{n}\left(y_{i}-\hat{y}_{i}\right)^{2}}{\sum_{i=1}^{n}\left(y_{i}-\bar{y}\right)^{2}}
\end{equation}
where $\hat{y}_i$ is the predicted value of the $i$th sample, $y_{i}$ is the actual value of the sample, $\bar{y}$ is the mean of all actual data and $n$ is the number of samples. The larger the $R^{2}$, the better the prediction. 

In this work, a grid search is applied to tune the hyperparameters in the CNN. A CNN consists of convolutional layers, pooling layers, and fully connected layers. In the convolutional layer, the hyperparameters include the number of filters, kernel size, and stride numbers. For the pooling layer, the hyperparameters include the filter size, padding, and stride numbers. We also need to decide the number of convolutional blocks, the number of fully connected layers, and the number of neurons per layer. For hyperparameter tuning, we search for the number of convolutional blocks and filter size with a pre-chosen kernel size of five and a stride size of one. The number of convolutional blocks varies from two to five and the filter size varies from 32 to 256. We also search for the number of hidden layers in the fully connected layers and the number of neurons per layer, with hidden layer numbers varying from one to five, and the number of neurons varying from 10 to 100 with a step of 10. The performance of each CNN structure is evaluated based on the $R^{2}$ score. 

\section{DATA AVAILABILITY}
The main data that support the findings of this study are presented in the paper and the Supplementary information. Extra data are available from the corresponding author upon reasonable request.

\section{ACKNOWLEDGEMENTS}
This work was supported by the National Science Foundation (NSF) under Grant No.~CMMI-1934367 and the Beijing Institute of Collaborative Innovation under Award No.~20183405. Jennifer A. Glerum and Jennifer Bennett acknowledge support by the US Army Research Laboratory under Grant No.~W911NF-19-2-0092. The SEM analysis work made use of the EPIC facility of NUANCE Center and the MatCI Facility of the Materials Research Center at Northwestern University, which was supported by NSF under Grant No.~ECCS-1542205 and DMR-1720139, the International Institute for Nanotechnology (IIN), the Keck Foundation, and the State of Illinois through the IIN.

\section{AUTHOR CONTRIBUTIONS}
L.F. conceptualized the study, developed the thermal model, designed the machine learning models, analyzed the results, wrote the manuscripts and supplementary information. L.C. contributed to discussion of data-driven models and results analysis. J.A.G. conducted the SEM microstructure characterization. J.B. performed the DED experiments and tensile tests. J.C. supervised the DED experiments. G.J.W. supervised the project, co-conceptualized the study, and co-wrote and edited the manuscript. All authors contributed to the discussion of the results and manuscript preparation. 

\section{COMPETING INTERESTS}
The authors declare no competing interests. 

\section{ADDITIONAL INFORMATION}
Supplementary information is available.







\bibliographystyle{elsarticle-num-names}
\bibliography{sample.bib}







\end{document}


\begin{frontmatter}


\title{Supplementary Information: Data-driven analysis of thermal simulations, microstructure and mechanical properties of Inconel 718 thin walls deposited by metal additive manufacturing }



\author[1]{Lichao Fang}
\author[1]{Lin Cheng}
\author[2]{Jennifer A. Glerum}
\author[1,3]{Jennifer Bennett}
\author[1]{Jian Cao}
\author[1]{Gregory J. Wagner\corref{cor1}}

\cortext[cor1]{Corresponding author}

\address[1]{Department of Mechanical Engineering, Northwestern University, Evanston, IL, 60208, USA}
\address[2]{Department of Materials Science and Engineering, Northwestern University, Evanston, IL, 60208, USA}
\address[3]{DMG MORI, Hoffman Estates, IL 60192, USA}

\end{frontmatter}

\renewcommand{\figurename}{Supplementary Figure}
\renewcommand{\tablename}{Supplementary Table}

This PDF file includes:

Supplementary Figures 1 to 5

Supplementary Tables 1 to 7

References (1 to 7) \newpage



\begin{table}
	\footnotesize
	\tabcolsep 0pt
	\begin{center}
		\def\temptablewidth{1\textwidth}
		\caption{Details of the simulation cases.}   
		\label{table2.0:cases}
		{\rule{\temptablewidth}{1pt}}
		\begin{tabular*}{\temptablewidth}{@{\extracolsep{\fill}}lll}
			Cases & Computation domain of thin wall & Dwell time\\
			\hline
			Case A & $80\; \mathrm{mm} \times 3\; \mathrm{mm} \times60$ mm & 0 s \\
			Case B & $120\; \mathrm{mm} \times 3\; \mathrm{mm} \times60$ mm & 0 s \\
			Case C & $120\; \mathrm{mm} \times 3\; \mathrm{mm} \times60$ mm & 5 s \\
		\end{tabular*}
		{\rule{\temptablewidth}{1pt}}
	\end{center}
\end{table}

\begin{table}
	\footnotesize
	\tabcolsep 0pt
	\begin{center}
		\def\temptablewidth{1\textwidth}
		\caption{The thermal-physical properties of Inconel 718 used in the model \cite{chen2011fibre,wolff2019experimentally,knapp2019experiments}.}   
		\label{table2.3:Tproperty}
		{\rule{\temptablewidth}{1pt}}
		\begin{tabular*}{\temptablewidth}{@{\extracolsep{\fill}}llll}
			Name & Symbol & Value & Unit\\
			\hline
			Solidus temperature & $T_s$ & 1533 & K \\
			Liquidus temperature & $T_l$ & 1609 & K \\
			Boiling point & $T_{b}$ & 3120 & K \\
			Latent heat of fusion & $L$ & 272 & $\mathrm{kJ}/\mathrm{kg} $ \\
			Latent heat of vaporization & $L_v$ & $2.224\times 10^{3}$ & $\mathrm{kJ}/\mathrm{kg}$ \\
			Density & $\rho$ & 8100 & $\mathrm{kg}/\mathrm{m}^{3}$\\
			Molar mass of the evaporating species & $m_{\mathrm{mol}}$ & $166.534\times 10^{-3}$ & $\mathrm{kg}/\mathrm{mol}$ \\
		\end{tabular*}
		{\rule{\temptablewidth}{1pt}}
	\end{center}
\end{table}

\begin{table}
	\footnotesize
	\tabcolsep 0pt
	\begin{center}
		\def\temptablewidth{1\textwidth}
		\caption{The thermal-physical properties of stainless steel 304 for substrate material \cite{wang2011numerical,handbook1990volume,carvalho2018explosive}.}   
		\label{table2.3:Tproperty_steel}
		{\rule{\temptablewidth}{1pt}}
		\begin{tabular*}{\temptablewidth}{@{\extracolsep{\fill}}llll}
			Name & Symbol & Value & Unit\\
			\hline
			Solidus temperature & $T_s$ & 1670 & K \\
			Liquidus temperature & $T_l$ & 1727 & K \\
			Latent heat of fusion & $L$ & 247 & $\mathrm{kJ}/\mathrm{kg} $ \\
			Density & $\rho$ & 8000 & $\mathrm{kg}/\mathrm{m}^{3}$\\
			Specific heat & $C_{p}$ & 500.0 &$\mathrm{J}/(\mathrm{kg}\cdot\mathrm{K})$ \\
			Thermal conductivity & $k$ & 16.2 & $\mathrm{W}/(\mathrm{m}\cdot\mathrm{K})$\\
		\end{tabular*}
		{\rule{\temptablewidth}{1pt}}
	\end{center}
\end{table}

\begin{table}
	\footnotesize
	\tabcolsep 0pt
	\begin{center}
		\def\temptablewidth{1\textwidth}
		\caption{Process parameters used in the model.}   
		\label{table2.3:process}
		{\rule{\temptablewidth}{1pt}}
		\begin{tabular*}{\temptablewidth}{@{\extracolsep{\fill}}llll}
			Name & Symbol & Value & Unit\\
			\hline
			Laser power & $P$ & 1800 & W \\
			Laser scan speed & $V$ & 16.7 & $\mathrm{mm}/\mathrm{s}$ \\
			Beam radius & $r_b$ & 1.5 & mm \\
			Layer thickness & $\Delta{z}$ & 0.5 & mm\\
			Ambient temperature &  $T_{\infty}$ & 295 & K\\
			Stefan–Boltzmann constant & $\sigma$ & $5.67 \times 10^{-14}$ & $\mathrm{W} / (\mathrm{mm}^{2}\cdot \mathrm{K}^{4})$ \\
			Emissivity & $\varepsilon$ & 0.45 & 1 \\
			Laser absorption efficiency & $\eta$ & 0.45 & 1 \\
			Ambient pressure & $P_{a}$ & $1.01\times 10^{5}$ & Pa \\
			Gas constant & $R_{\mathrm{gas}}$ & 8.314 & $\mathrm{J}/(\mathrm{K}\cdot\mathrm{mol})$ \\
		\end{tabular*}
		{\rule{\temptablewidth}{1pt}}
	\end{center}
\end{table}


\begin{table}
	\scriptsize
	\tabcolsep 0pt
	\begin{center}
		\def\temptablewidth{1\textwidth}
		\caption{Cooling rate, microstructure features, and mechanical properties for different wall build cases. Here $\dot{T}$ is the cooling rate, $\lambda_1$ is the primary dendrite arm spacing, $\phi_{\mathrm{Laves}}$ is the volume fraction of Laves phases, UTS is the ultimate tensile strength, $\sigma_Y$ is the yield stress, $\sigma_F$ is the failure stress, and $E$ is the modulus.} 
		\label{table5.2:PSPwall}
		{\rule{\temptablewidth}{1pt}}
		\resizebox{\textwidth}{!}{
			\begin{tabular*}{\temptablewidth}{@{\extracolsep{\fill}}cccccccc}
				& $\dot{T}$ (K/s) & $\lambda_1 (\mathrm{\mu m})$  & $\phi_{\mathrm{Laves}}$ (\%) & UTS (MPa) & $\sigma_Y (\mathrm{MPa})$ & $\sigma_F (\mathrm{MPa})$ & $E (\mathrm{MPa})$ \\
				\hline
				Case A & 19.58&	16.89&	2.59&	649.41&	388.27&	520.28&	12641.4\\
				Case B & 27.30&	11.92&	3.26&	751.99&	496.64&	596.64&	13366.3\\
				Case C & 43.05&	7.98&	3.34&	782.44&	523.60&	635.25&	14125.8\\
				
		\end{tabular*}}
		{\rule{\temptablewidth}{1pt}}
	\end{center}
\end{table}
\begin{table}
	\scriptsize
	\tabcolsep 0pt
	\begin{center}
		\def\temptablewidth{1\textwidth}
		\caption{Cooling rate, microstructure features, and mechanical properties for different locations in the Case B wall. Here, $\dot{T}$ is the cooling rate, $\lambda_1$ is the primary dendrite arm spacing, $\phi_{Laves}$ is the volume fraction of Laves phases, UTS is the ultimate tensile strength, $\sigma_Y$ is the yield stress, $\sigma_F$ is the failure stress, and $E$ is the modulus.} 
		\label{table5.3:PSP8}
		{\rule{\temptablewidth}{1pt}}
		\resizebox{\textwidth}{!}{
			\begin{tabular*}{\temptablewidth}{@{\extracolsep{\fill}}cccccccc}
	        &  $\dot{T}$ (K/s) & $\lambda_1 (\mathrm{\mu m})$  & $\phi_{Laves}$ (\%) & UTS (MPa) & $\sigma_Y (\mathrm{MPa})$ & $\sigma_F (\mathrm{MPa})$ & $E (\mathrm{MPa})$ \\
			\hline
			Point 1 (top) &27.24&	11.88&	2.23&	720.52&	480.60&	572.18	&15667.37\\
			Point 8 (middle) & 26.44&	14.98&	2.76&	776.54&	568.05&	733.34&	18698.09\\
			Point 12 (bottom) &25.09&	12.87&	2.38&	749.28&	540.80	&608.51	&17940.91\\			
		\end{tabular*}}
		{\rule{\temptablewidth}{1pt}}
	\end{center}	
\end{table}
\begin{figure}
	\centering\includegraphics[width=3in]{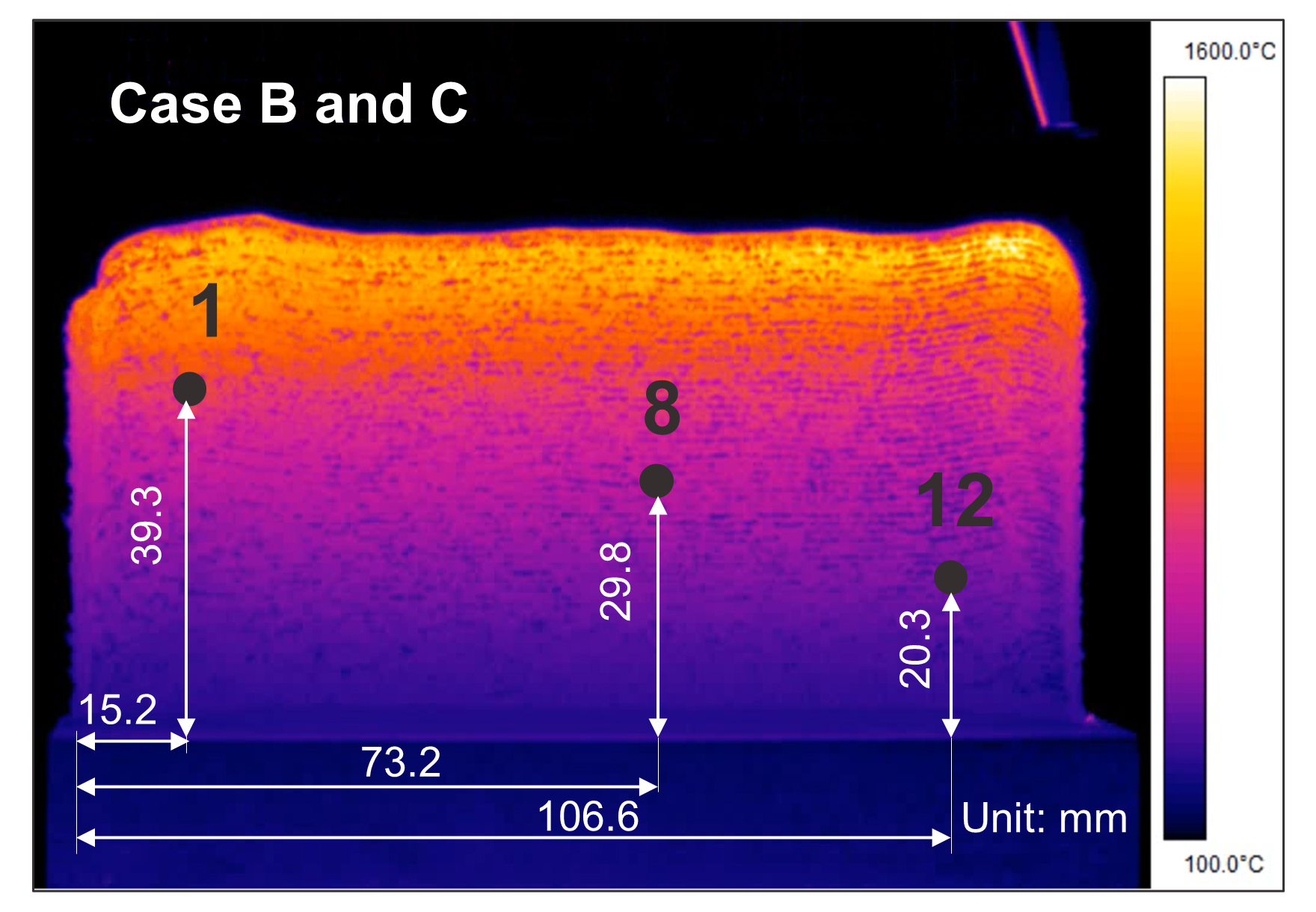}
	\caption{Sample point locations for microstructure measurements in Case B.}
	\label{fig5.3:IR8}
\end{figure}

\begin{table}
\footnotesize
\tabcolsep 0pt
\begin{center}
\def\temptablewidth{1\textwidth}
\caption{The temperature dependent thermal properties of material Inconel 718 \cite{diaz2014numerical}. Properties at intermediate temperatures are calculated by linear interpolation; properties at temperatures above 1573K are assumed constant at the upper limit.}   
\label{table2.1:Tthermal718}
{\rule{\temptablewidth}{1pt}}
\begin{tabular*}
{\temptablewidth}{@{\extracolsep{\fill}}ccc}
$T$(K) & $k (\mathrm{W}/(\mathrm{m}\cdot\mathrm{K}))$ & $C_{p}(\mathrm{J}/(\mathrm{kg}\cdot\mathrm{K}))$  \\
\hline
293.0 &	10.31&	362\\
373.0&	11.88&	378\\
473.0&	13.6&	400\\
673.0&	16.6&	412\\
873.0&	20.1&	460\\
1073.0&	26.3&	544\\
1573.0&	30.75&	583\\
\end{tabular*}
{\rule{\temptablewidth}{1pt}}
\end{center}
\end{table}

\begin{figure}
	\centering
	\includegraphics[width=3in]{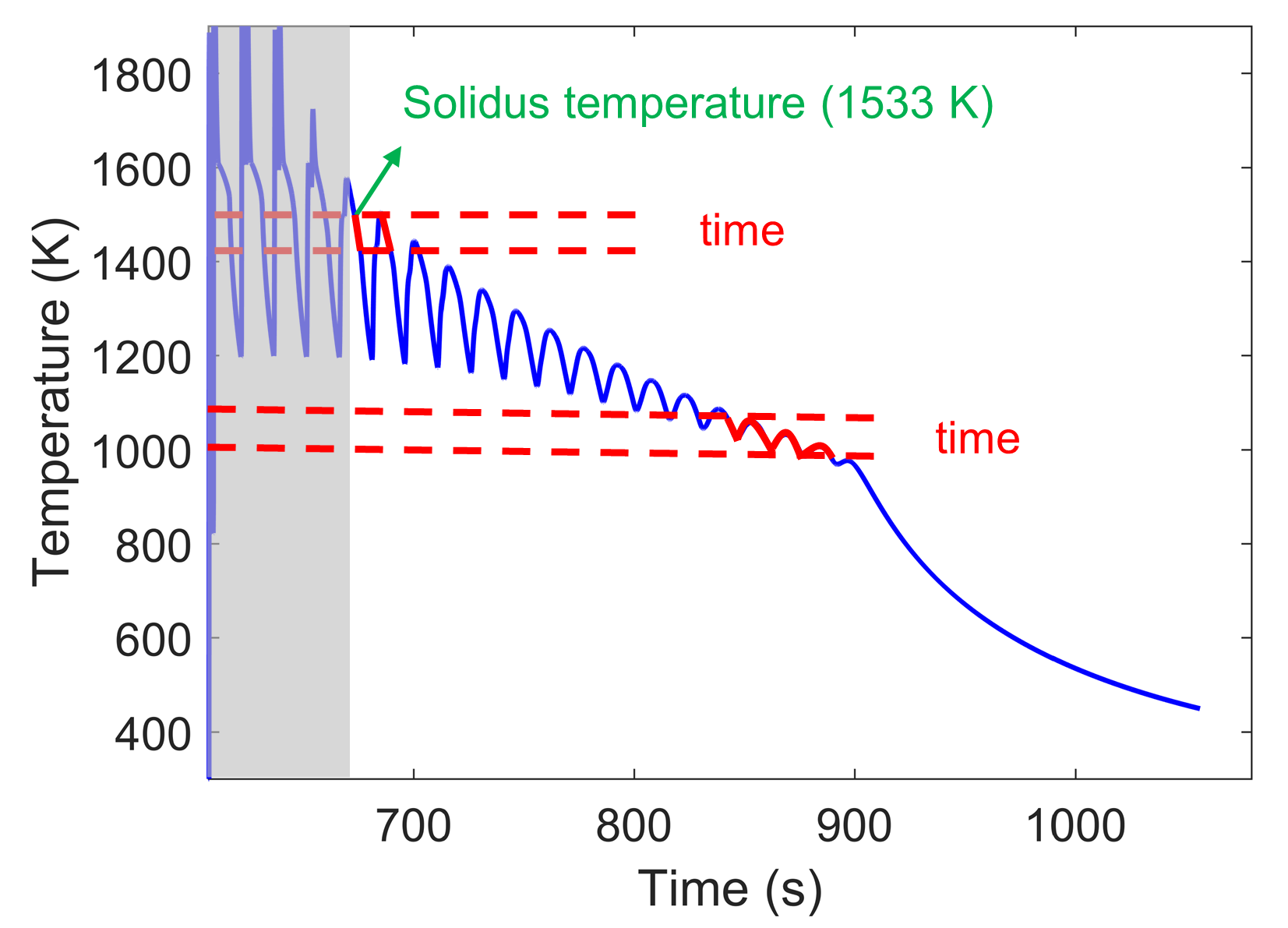}
	\caption{Post-processing to extract a reduced set of thermal history feature data. The thermal history is divided into several sections with 25 K intervals, for example the range marked by dotted red lines, and cumulative time spent during each temperature interval is calculated. Only the thermal history after the last crossing of the solidus temperature (the white region) is considered.}
	\label{fig4.2:Tinternal}
\end{figure}

\begin{figure}
	\centering
	\includegraphics[width=3in]{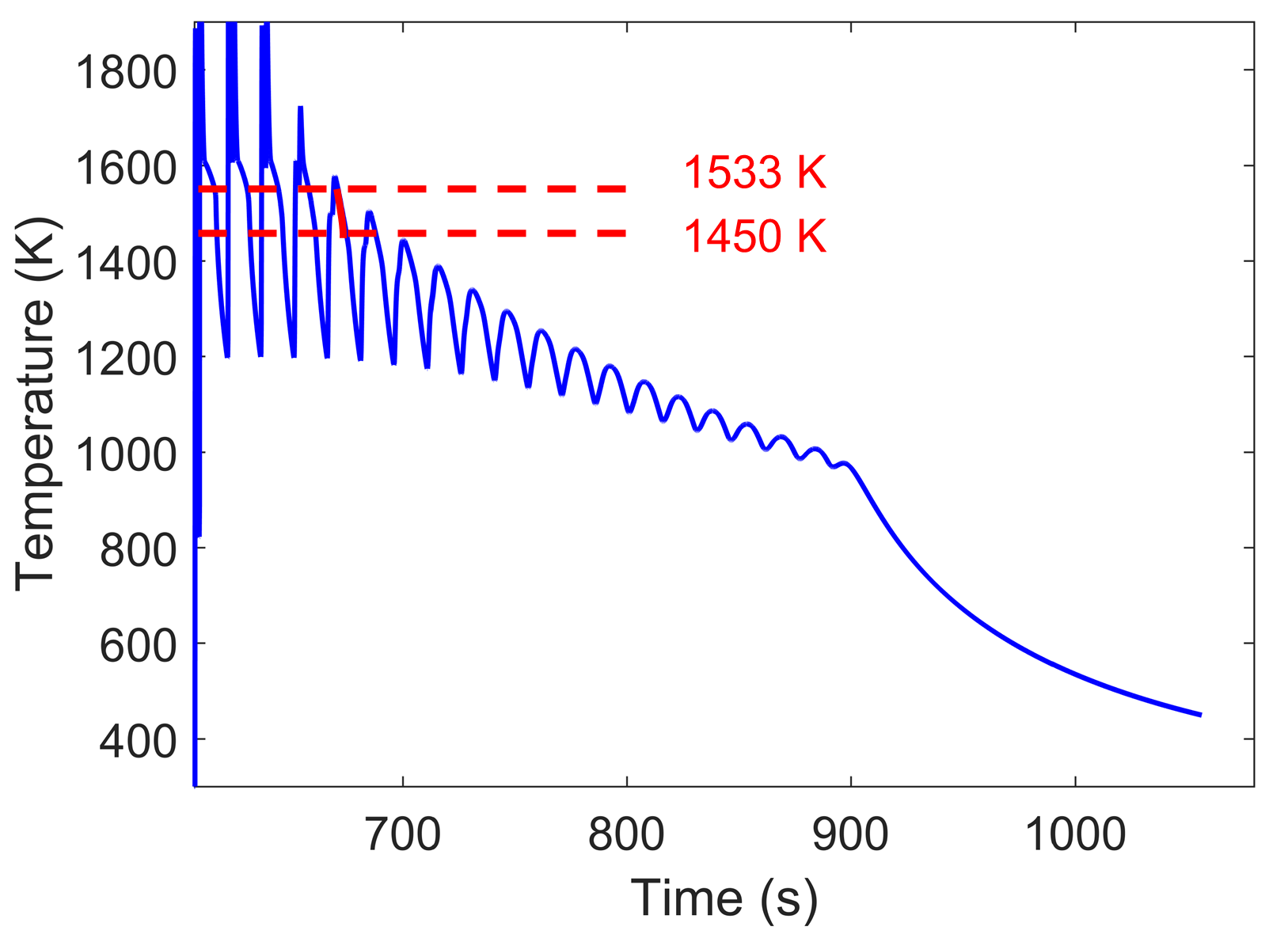}
	\caption{Cooling rate is calculated as the average slope of the temperature history between the last crossing of the solidus temperature (1533 K) and an arbitrarily chosen temperature of 1450 K.}
	\label{fig4.3:coolingrate}
\end{figure}
\label{sec:CoolingRate}

\begin{figure}
	\centering
	\includegraphics[height=1.4in]{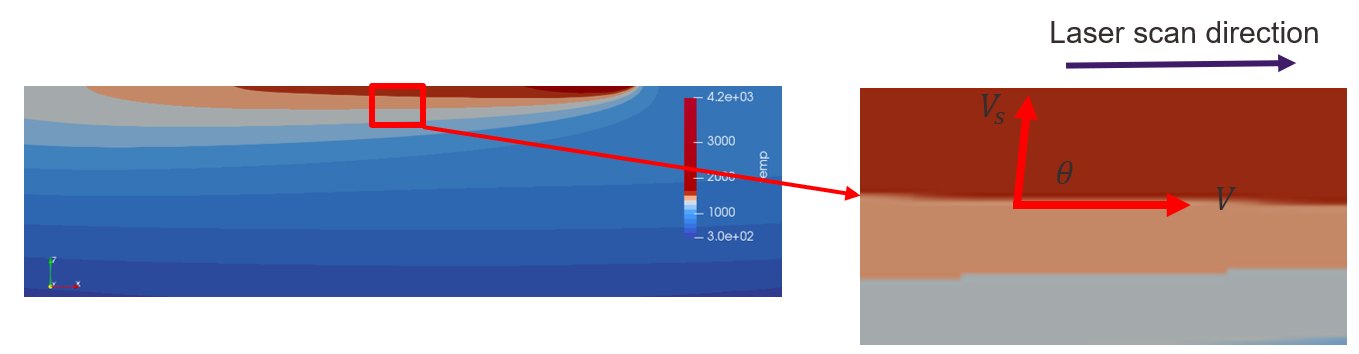}
	\caption{Schematic of calculating solidification rate in the 2D section of melt pool centerline in 3D DED deposition simulation. The liquid-solid interface is divided by the red and orange color. The solidification rate $V_s$ is calculated by $V_{s}=V\cos\theta$, where $V$ is the laser scan speed and $\theta$ is the angle between laser scan direction and the direction of the normal vector to the liquid-solid interface. }
	\label{fig4.4:mesh_angle}
\end{figure}
\begin{figure}
	\centering
	\includegraphics[width=3in]{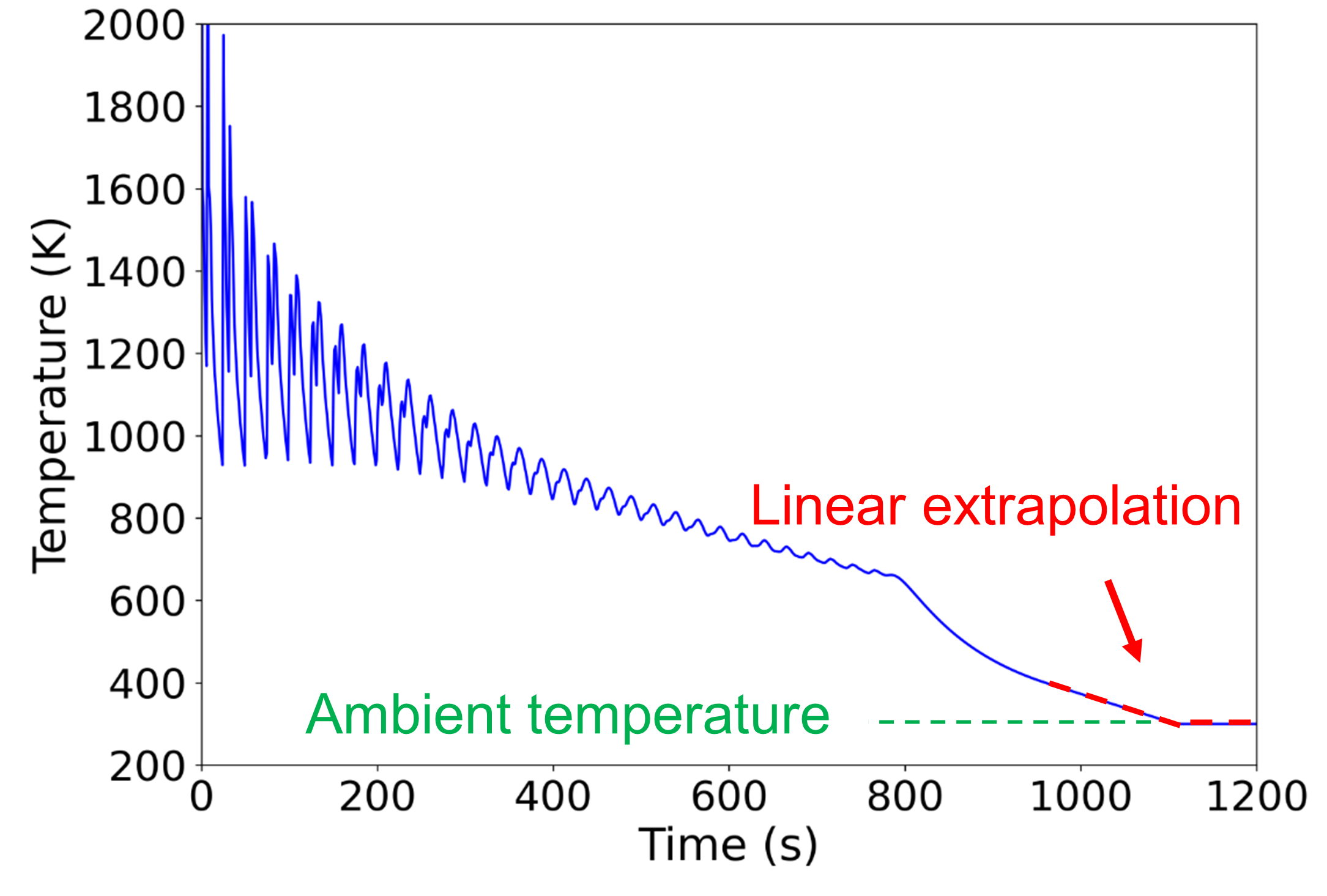}
	\caption{Data preparation for input to the CNN. Linear extrapolation (red dashed line) is applied to the end of original thermal history, down to a constant ambient temperature, to ensure same input data size for all thermal histories.}
	\label{fig5.7:CNN_input_prep}
\end{figure}

\newpage




\FloatBarrier


\bibliographystyle{elsarticle-num-names}
\bibliography{sample.bib}





